\documentclass[twocolumn]{aastex62}

\usepackage{multirow}
\usepackage{lineno}
\usepackage{graphicx}  
\usepackage{dcolumn}   
\usepackage{bm}        
\usepackage{amssymb}   
\usepackage{amsmath}
\usepackage{longtable}
\usepackage{lipsum}
\usepackage[bottom]{footmisc}

\shorttitle{Co-production of $r$-process elements via fission}
\shortauthors{Vassh et al.}

\begin{document}

\title{Co-production of Light and Heavy $r$-process Elements via Fission Deposition}

\correspondingauthor{Nicole Vassh}
\email{nvassh@nd.edu}

\author{Nicole Vassh}
\affiliation{University of Notre Dame, Notre Dame, IN 46556, USA}

\author{Matthew R. Mumpower}
\affiliation{Theoretical Division, Los Alamos National Lab, Los Alamos, NM 87545, USA}
\affiliation{Joint Institute for Nuclear Astrophysics - Center for the Evolution of the Elements, USA}

\author{Gail C. McLaughlin}
\affiliation{North Carolina State University, Raleigh, NC 27695, USA}

\author{Trevor M. Sprouse}
\affiliation{University of Notre Dame, Notre Dame, Indiana 46556, USA}

\author{Rebecca Surman}
\affiliation{University of Notre Dame, Notre Dame, Indiana 46556, USA}
\affiliation{Joint Institute for Nuclear Astrophysics - Center for the Evolution of the Elements, USA}



\begin{abstract}
We apply for the first time fission yields determined across the chart of nuclides from the macroscopic-microscopic theory of the Finite Range Liquid Drop Model to simulations of rapid neutron capture ($r$-process) nucleosynthesis. With the fission rates and yields derived within the same theoretical framework utilized for other relevant nuclear data, our results represent an important step toward self-consistent applications of macroscopic-microscopic models in $r$-process calculations. The yields from this model are wide for nuclei with extreme neutron excess. We show that these wide distributions of neutron-rich nuclei, and particularly the asymmetric yields for key species that fission at late times in the $r$ process, can contribute significantly to the abundances of the lighter heavy elements, specifically the light precious metals palladium and silver. Since these asymmetric yields correspondingly also deposit into the lanthanide region, we consider the possible evidence for co-production by comparing our nucleosynthesis results directly with the trends in the elemental ratios of metal-poor stars rich in $r$-process material. We show that for $r$-process enhanced stars palladium over europium and silver over europium display mostly flat trends suggestive of co-production and compare to the lanthanum over europium trend which is often used to justify robustness arguments in the lanthanide region. We find that such robustness arguments may be extendable down to palladium and heavier and demonstrate that fission deposition is a mechanism by which such a universality or robustness can be achieved. 
\end{abstract}

\keywords{Processes: nucleosynthesis --- stars: abundances}



\section{Introduction}\label{sec:intro} 
 
An understanding of the observed solar abundances for elements heavier than iron requires the disentangling of contributions from several astrophysical processes. After subtracting off nuclei on the proton-rich side of stability as well as contributions from the slow neutron capture process ($s$ process), one is left with what is often taken to be the contribution from the rapid neutron capture process, that is the $r$-process residuals. However such abundances are not necessarily representative of solely $r$-process nucleosynthetic outcomes as all other potential astrophysical contributions are hidden within these residuals. 
 
In order to accommodate the solar $r$-process residuals of both the lighter heavy elements (between the first and second $r$-process peaks at $A\sim80$ and $A\sim130$, respectively) as well as those of the heavier nuclei such as platinum and uranium, several astrophysical processes are likely needed (e.g. \cite{ThielemannEtAl2011}). Since nature offers many possible ways to synthesize the lighter heavy elements, the story of how such elements came to populate the cosmos is likely to be rich and complex. For instance, electron capture supernovae are among the possible sites of interest \citep{Wanajo2011} and, depending on the progenitor, some core-collapse supernova simulations have suggested synthesis up to silver to be possible (e.g. \cite{ArconesMontes,Bliss}). Additionally the $\nu$p process can proceed up to $A\sim100$ in some conditions (e.g. \cite{ThielemannEtAl2010}). An intermediate neutron capture process ($i$ process) taking place in rapidly accreting white dwarfs (e.g. \cite{Coteiprocess,DenissenkovEtA2018}) or in neutrino dominated explosions enhanced by magnetic fields \citep{Nishimura2017} is another possible source.

The electromagnetic counterpart to the neutron star merger event GW170817 has suggested that such events also contribute to lighter heavy elements since observations saw an early `blue' kilonova component as well as a late `red' kilonova associated with high opacity lanthanide elements (e.g. \cite{Cowperthwaite2017,AbbottGW170817,Villar}). This could be explained via separate contributions by a `weak' $r$ process, which terminates at or before the production of second peak nuclei, and a `strong' or `main' $r$ process that populates past the second peak elements into the lanthanide region and beyond \citep{MetzgerKilonova2010,Kasen,RosswogGW170817,EvenCompositionKilo}. Such a result could be achieved by a two-component merger model consisting of very neutron-rich dynamical ejecta to produce the main $r$ process as well as a later accretion disk wind that can fill in the lighter heavy elements \citep{CoteGW170817,MillerBlueKilo}. It should be noted that some simulations show that dynamical ejecta alone can produce lighter heavy elements along with a strong $r$ process by having a fraction of their ejecta mass undergo solely a weak $r$ process (e.g. \cite{Radice18}).

To help disentangle the possible contributions of various nucleosynthesis sites in our Galaxy, metal-poor stars that are enriched in $r$-process elements such as europium can provide crucial insights. Since such stars are either old or born in pristine environments, they are thought to probe one to few $r$-process events and therefore can provide a less convoluted picture of the details of the astrophysical $r$ process than can be understood from solar abundances. An intriguing feature that emerges when comparing the relative abundances of metal-poor, r-rich stars is the so-called universality of the pattern for elements with $Z\ge56$ which includes the lanthanide elements \citep{Sneden}. The stability of the abundance patterns from star to star is often also pointed to as a argument for the $r$ process to be robust, that is, to always produce the similar elemental ratios. Why such a universality is found in nature when nucleosynthetic outcomes from various astrophysical simulations show dependences on simulation conditions such as progenitor mass remains unknown. One suggestion for a mechanism by which universality can be achieved is via a fission cycling $r$ process where final abundances are largely set by the fission fragment distributions of neutron-rich nuclei \citep{Beun08,Goriely+11,Oleg,GorielyGEF,Mendoza-Temis+15}. 

In this work we revisit the question of universality with the discussion extended to consider a subset of the lighter heavy elements: the light precious metals ruthenium (Ru), rhodium (Rh), palladium (Pd), and silver (Ag). These elements have previously been argued to be dominantly synthesized by a light element primary process (LEPP) (e.g. \cite{MontesApJ07,Montes07Conf,Travaglio}). For instance \cite{MontesApJ07} used trends in the elemental ratios observed in metal-poor stars of [X/Eu], where X is various lighter heavy elements such as silver and Eu is europium, to explore the conditions consistent with a LEPP that ranged from $s$-process to $r$-process type neutron densities (in which case the LEPP is essentially equivalent to a weak $r$ process). Following this a larger observational data set for metal-poor stars also reported trends in palladium and silver over europium that indicate that these light precious metals can be synthesized independently from a main $r$ process \citep{Hansen12}. In this work we focus on observations of the subset of metal-poor stars that show enhanced abundances of main $r$-process elements (the so-called r-I and r-II stars \citep{AbohalimaFrebel}) in order to consider whether a robust $r$ process, as can be produced in merger dynamical ejecta, can contribute to the light precious metals via a previously unexplored mechanism: late-time fission deposition. 

Examining the effects of fission in astrophysical environments requires knowledge of fission properties for hundreds of nuclei on the neutron-rich side of stability, about which little is experimentally known. Such a lack of available nuclear data for neutron-rich nuclei is not only a problem encountered with fission but with every reaction and decay channel that is involved in the $r$ process. Thus dealing with the nuclear data uncertainties affecting predictions for the $r$-process outcome of astrophysical events is a key component in developing a deeper understanding of how the heavy elements observed in the Galaxy came to be populated. Since presently $r$-process calculations must rely heavily on theoretical descriptions that can vary widely, an important aspect of reducing calculation uncertainties is to push toward consistent treatments of the theoretical data so that features seen in predicted abundances are not in fact an artifact of a mismatch between the properties of nuclei assumed for a given reaction channel, such as neutron capture, and the properties assumed for the same nuclei in the data applied for other channels, such as $\beta$-decay.

A piece of nuclear data of particular importance in a fission cycling $r$ process is the fission fragment treatment \citep{Goriely+13,GorielyGEF,GorielyGMPGEF,Panov,Kodama,Eichler15,Eichler16}. Phenomenological descriptions, such as ABLA \citep{ABLA91,ABLA07}, Wahl \citep{Wahl}, and GEF \citep{GEF}, are presently the standard in $r$-process calculations \citep{Mendoza-Temis+15,Mendoza-Temis+16,RobertsWahl,GorielyGEF,GorielyGMPGEF,VasshGEF2019}. Fission theories based on macroscopic-microscopic models or density functional theory have begun advancing into the neutron-rich regions but until recently no theoretical predictions were available across the broad range of neutron-rich nuclei accessed during a fission cycling $r$ process. We investigate for the first time the astrophysical implications from applying the new Finite Range Liquid Drop Model (FRLDM) fission yields recently developed at Los Alamos National Laboratory \citep{Mumpower+19} in neutron-rich merger ejecta. Although reaction and decay rates derived within a macroscopic-microscopic framework such as the Finite Range Drop Model (FRDM) \citep{FRDM2012} are commonly applied in nucleosynthesis calculations, this work is the first to apply fission yields from a related macroscopic-microscopic model in the $r$ process. Additionally since in this work we apply fission rates derived within the same theoretical framework as these yields, our calculations take an important step toward self-consistent treatments of fission in the $r$ process.

The paper is organized as follows: in Section~\ref{sec:rpimpact} we provide a brief overview of the FRLDM fission yields and demonstrate the $r$-process impact of these fission yields in conditions that could be found in merger ejecta. In Section~\ref{sec:twocomp} we explore the implications of light heavy element contributions from fission when a two-component merger model of accretion disk wind plus dynamical ejecta is considered given a simulation of dynamical ejecta in which all conditions present find fission to occur robustly. In Section~\ref{sec:radicedyn} we consider a simulation of merger dynamical ejecta that produces both a weak and main $r$ process by having a broad range of conditions and demonstrate the contribution from fission in such a scenario. In Section~\ref{sec:rstars} we investigate observational hints for the co-production of the light precious metals and heavier $r$-process nuclei, such as the lanthanides, by considering elemental ratios seen in metal-poor, $r$-process enhanced stars and compare to our nucleosynthetic yields. We conclude in Section~\ref{sec:conclude}.

\section{FRLDM fission yields in merger ejecta}\label{sec:rpimpact}

In this work we use primary fission fragment yields from the FRLDM model, as detailed in \cite{Mumpower+19}. 
These yields are generated assuming strongly damped nuclear dynamics that leads to the description of fission process via Brownian shape motion across nuclear potential-energy surfaces. 
We assume that the excitation energy of each fissioning system is just above the height of the largest fission barrier. 
We further assume in using the primary fragment yield that the prompt neutron emission associated with fission plays a minor role in the synthesis of elements. 
Both of these approximations have been shown to be suitable for applications of $r$-process nucleosynthesis \citep{VasshGEF2019}. 

\begin{figure}
\begin{center}
\includegraphics[scale=0.475]{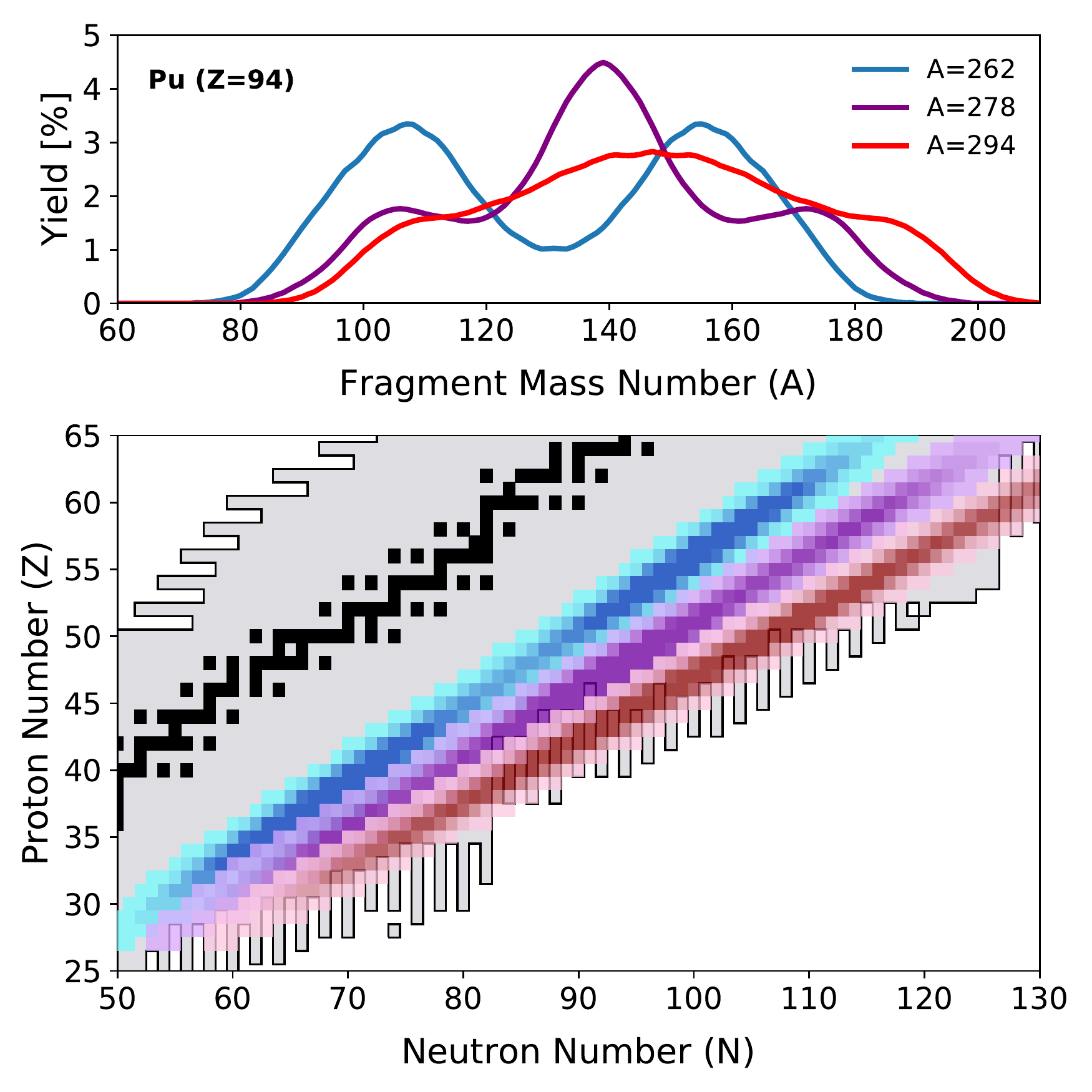}
\end{center}
\caption{The FRLDM fission yields for three neutron-rich isotopes in the plutonium isotopic chain shown as a function of fragment mass number (upper) as well as in the NZ plane (lower). Gray shows nuclei within the FRDM2012 dripline.}
\label{fig:FRLDMyields}
\end{figure}

We demonstrate the tendency for FRLDM yields to deposit a broad range of fission product species in Figure~\ref{fig:FRLDMyields} by showing isotopes of plutonium of increasing neutron richness. With this yield model, the fission product mass numbers show the widest range for the most neutron-rich fissioning species past the $N=184$ shell closure, such as plutonium-294, which sees the production of $A\sim110$ nuclei almost as likely as the $A\sim150$ product nuclei near the symmetric peak. This heaviest isotope of plutonium in Fig.~\ref{fig:FRLDMyields} demonstrates that the yields of very heavy neutron-rich nuclei can deposit daughter products outside the neutron dripline. When this occurs within an $r$-process calculation, we assume such a species emits neutrons instantaneously until reaching an isotope with a positive one-neutron separation energy. Such neutron-rich fission products can further contribute to the free neutrons available for capture by undergoing $\beta$-delayed neutron emission as discussed in \cite{Mendoza-Temis+15}. 

The yield trend of the FRLDM model when going from the most neutron-rich fissioning species to less neutron-rich isotopes toward stability transitions from symmetric to asymmetric, as shown by plutonium-262 in Fig.~\ref{fig:FRLDMyields}. Such asymmetric yields give fission products more concentrated near $A\sim110$ and $A\sim155$ but still show broad deposition. Thus a nonnegligible amount of fission deposition occurs at neutron numbers lower than the $N=82$ shell closure for many neutron-rich isotopes of importance in the $r$ process.

For investigating the nucleosynthesis impact of this yield model, we use the network Portable Routines for Integrated nucleoSynthesis Modeling (PRISM) developed jointly at the University of Notre Dame and Los Alamos National Laboratory \citep{BDFrp,VasshGEF2019}. PRISM permits a straightforward implementation of mass model-dependent nucleosynthesis rates due to its flexibility with nuclear data inputs. For the masses of neutron-rich nuclei, we apply the Finite Range Drop Model (FRDM2012). Where available we use experimental masses \citep{AME2016} as well as experimentally established half-lives and branching ratios from NUBASE \citep{NUBASE2016}. For theoretical $\alpha$-decay rates we use the well-established Viola--Seaborg formula \citep{VS1966} where we apply a least-squares fit to NUBASE2016 half-life data that takes into account the reported experimental uncertainties when optimizing coefficients as in \cite{VasshGEF2019}. We use neutron capture, $\beta$-decay, neutron-induced fission, and $\beta$-delayed fission rates as in \cite{Kawano2008,Kawano2016,Kawano+17,Mumpower+16,BDFrp,MollerQRPA,VasshGEF2019}, with all rates determined from the same model masses as in \cite{Mumpower+15} and updated to be self-consistent with the fission barrier heights of a given model. For spontaneous fission we apply a parameterized prescription with a simple dependence on barrier height as in \cite{Karpov,Zagrebaev} (we note that since this and other treatments \citep{Samuel18} often find spontaneous fission to not significantly influence nucleosynthetic abundances, in this work we focus on the impact of neutron-induced and $\beta$-delayed fission). Here we employ FRLDM fission barriers in order to be consistent with the FRLDM inputs used to determine the fission yields. Therefore with the same fission barriers used to determine the fission yields and rates of all fission reaction and decay channels, our calculations are an important step toward consistent macroscopic-microscopic treatments of fission for neutron-rich nuclear data.

\begin{figure}
\begin{center}
\includegraphics[scale=0.49]{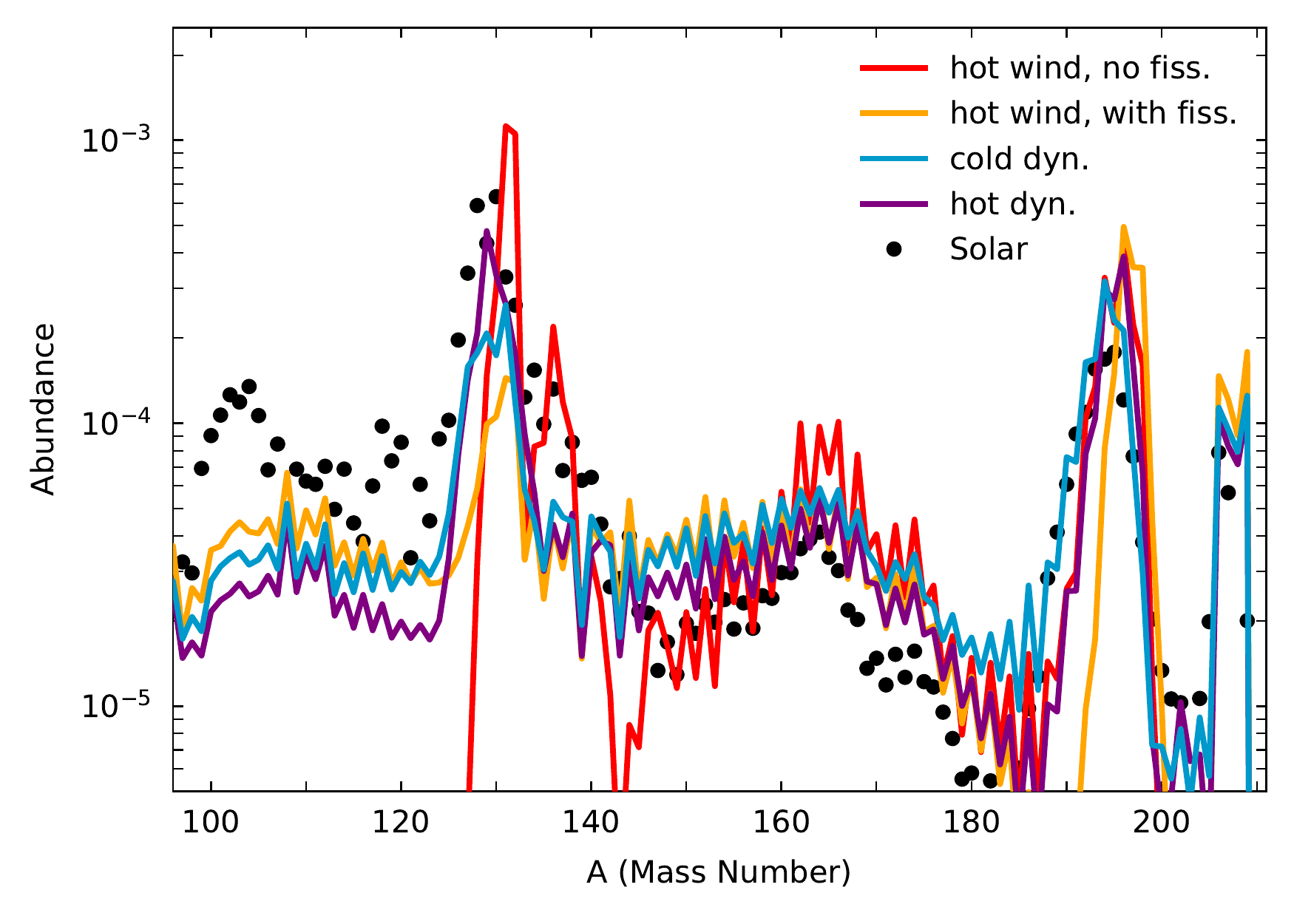}
\end{center}
\caption{A comparison of the $r$-process abundances using FRLDM fission yields in hot (purple) and cold (light blue) dynamical ejecta conditions. Here the cold case represents robust fission cycling conditions while the hot case does not show strong cycling behavior. We compare to conditions representative of a hot accretion disk wind that does not fission cycle at $Y_e=0.2$ (red) as well as the same wind conditions at $Y_e=0.15$ (orange) that will populate fissioning nuclei. Here we consider single trajectories rather than a mass-weighted average to demonstrate the fission deposition influence in distinct conditions. The scaled solar data is that of Goriely \citep{goriely99,Arnould07}.}
\label{fig:compconditions}
\end{figure}

The nucleosynthetic outcome in conditions that host fission depends strongly on the fission yields. With the nuclear inputs fixed, variances in $r$-process abundances arise solely from the range in astrophysical trajectories predicted by simulations. We first make use of the merger dynamical ejecta simulations from \cite{Rosswog} in order to investigate conditions that robustly produce fissioning species. We consider trajectories from a 1.2--1.4 $M_{\odot}$ neutron star merger, all of which represent very neutron-rich ($Y_e \lesssim 0.05$) dynamical ejecta, and refer to them by their original number labeling (1--30) in order to permit direct comparisons with our results. Figure~\ref{fig:compconditions} shows the results using FRLDM yields given two distinct tracers: one a `cold' tidal tail ejecta mass element (trajectory 1) and one a `hot' dynamical condition (trajectory 22) that has experienced more shock heating than the tail. Here we take the reheating efficiency to be 10$\%$ and apply the term `cold' to an astrophysical trajectory for which $\beta$-decay is the primary channel in competition with neutron capture (rather than photodissociation), while the term `hot' implies conditions that support an extended (n,$\gamma$)$\rightleftharpoons$($\gamma$,n) equilibrium. The cold case explored here represents robust fission cycling conditions while the hot case does not show strong cycling behavior.

In the cold conditions of traj.$\,$1, the majority of material gets pushed out of the light precious metal region up to higher mass numbers, eventually accessing the neutron-rich, heavy nuclei around $A\sim295$ with very wide fission yields. Since the second peak is largely absent when fission cycling begins, and this yield model does not concentrate deposition near $A\sim130$, the second $r$-process peak is underproduced in such conditions. In contrast, in the hot conditions of traj.$\,$22, the equilibrium path maintains an abundance of nuclei at the $N=82$ shell closure throughout the calculation. Such conditions never reach the nuclei with the widest yields and fission deposition is mostly concentrated near $A\sim139$ as well as $A\sim110,155$. Given the variances seen for dynamical ejecta across merger simulations, it is difficult to say exactly how much cold versus hot conditions are present in the ejecta. Should the ejecta have a significant amount of cold material, this yield model predicts an underproduction of the second $r$-process peak, which could suggest this abundance feature to be due to an $r$-process source other than dynamical ejecta. Although the hot and cold cases show pronounced differences in the second peak, with this yield model the light precious metals, as well as the lanthanides, are robustly produced in both types of conditions. 

For comparison in Fig.~\ref{fig:compconditions} we also show results given the parameterized conditions of a low entropy accretion disk wind with $Y_e=0.2$ (as considered in \cite{OrfordVassh2018}), which produces a main $r$ process but does not synthesize fissioning nuclei. Such conditions robustly produce lanthanides but fail to also populate the light precious metals. Astrophysical sites that do not host fission only see co-production of the light precious metals and heavier $r$-process elements by having a distribution of conditions present that separately contribute to these regions and are therefore subject to potentially larger variances in the ratios of the light precious metals to heavier nuclei. We explore this point in Sections \ref{sec:twocomp},\ref{sec:radicedyn}, and \ref{sec:rstars}. We also demonstrate in Fig.~\ref{fig:compconditions} that the same hot accretion disk wind conditions are capable of populating the light precious metals with $Y_e=0.15$. Therefore it is not solely very low $Y_e$ ($<0.05$) conditions, such as the dynamical ejecta considered here, that are capable of co-producing the light precious metals and lanthanides, rather all that is required is that fissioning nuclei participate during $r$-process nucleosynthesis.

\begin{figure}
\begin{center}
\includegraphics[scale=0.575]{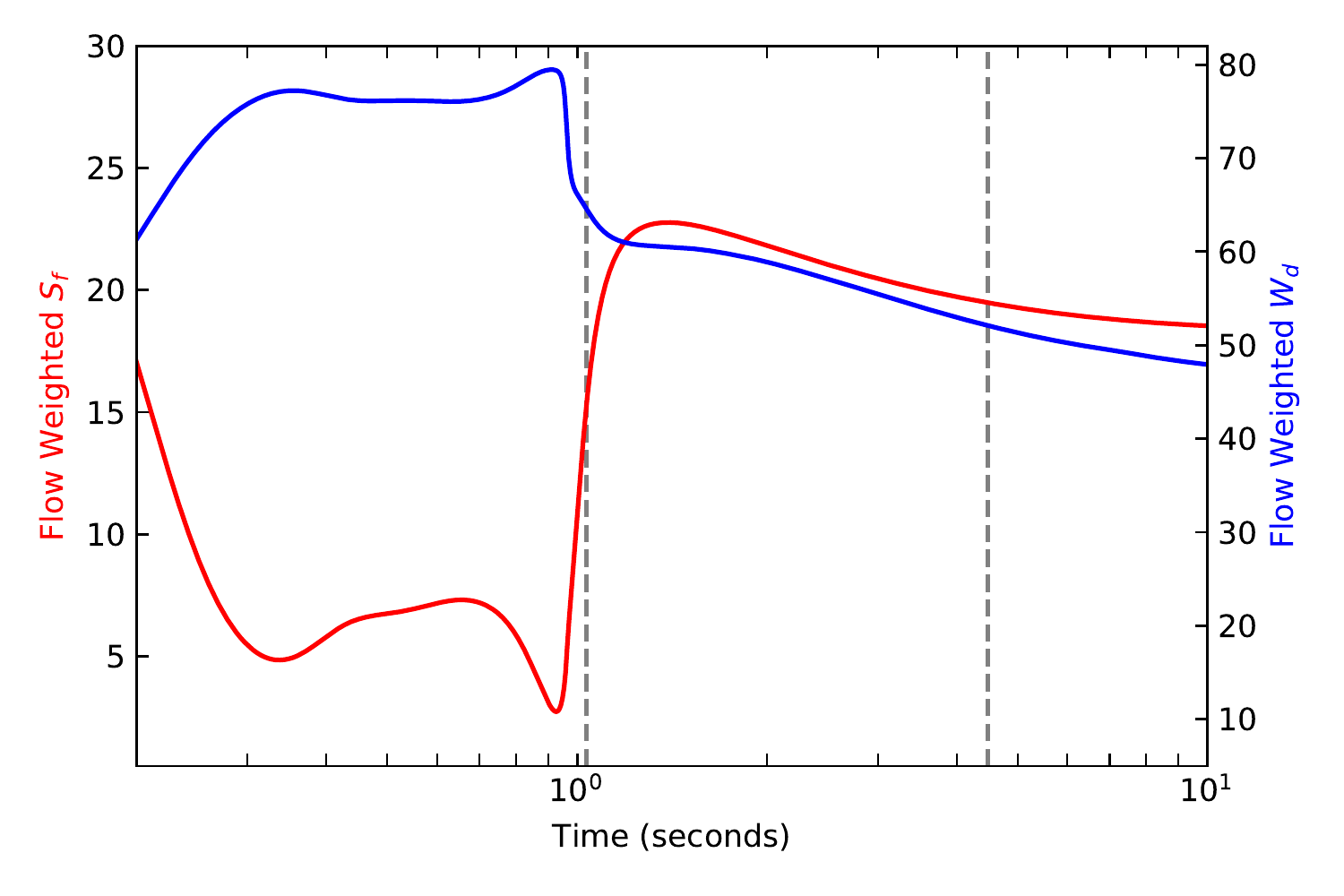}
\end{center}
\caption{Flow weighted fission yield metrics (normalized by the total flow at a given time step) as a function of time for the symmetric factor $S_{\rm{f}} = \left|A_{max}-A_f/2\right|$ (red, left axis) and the overall yield distribution width (blue, right axis). The vertical dashed gray lines denote the late and early times considered in more detail in Figure~\ref{fig:yielddeposition}.}
\label{fig:yieldmetrics}
\end{figure}

To further quantify how fission deposition continuously evolves during the $r$ process with the FRLDM yield model, we consider fission flow (rate multiplied by abundance) weighted values for the fission yield metrics introduced in \cite{Mumpower+19}, normalized by the total fission flow at a given time step. In Figure~\ref{fig:yieldmetrics}, we show the evolution of the symmetric factor, $S_{\rm{f}} = \left|A_{max}-A_{f/2}\right|$, where $A_{max}$ is the mass number at the maximum of the fission yield and $A_f$ is the mass number of the fissioning nucleus. Lower values of $S_f$ indicate the fission yield distribution to be symmetric, with typical values for the asymmetric distributions of experimentally probed actinides being around $S_{\rm{f}} \sim 20$. We also consider the overall width of the distribution, $W_{\rm{d}}$, defined to be the range in mass number that sees yield contributions above $1\%$. Typical values for the major actinides are around $W_{\rm{d}}\sim40$. Here we examine the cold dynamical ejecta conditions of traj.$\,$1 from Fig.~\ref{fig:compconditions} that first accesses fissioning nuclei which are symmetric, with low $S_{\rm{f}}$ values, and very wide distributions with $W_{\rm{d}}\sim80$. Then, due to fission cycling, a few asymmetric yields along with mostly symmetric yields are accessed around $0.7$ seconds followed by a re-emergence of primarily symmetric yield contributions after a bulk of fission cycled material makes its way back to the most neutron-rich regions just before the decay back to stability begins to dominate the $r$ process. After this time mostly asymmetric yields are accessed, but although the overall distribution width decreases, deposition still occurs over a range of $\sim60$ mass numbers.

\begin{figure}%
\begin{center}
\includegraphics[width=8.8cm]{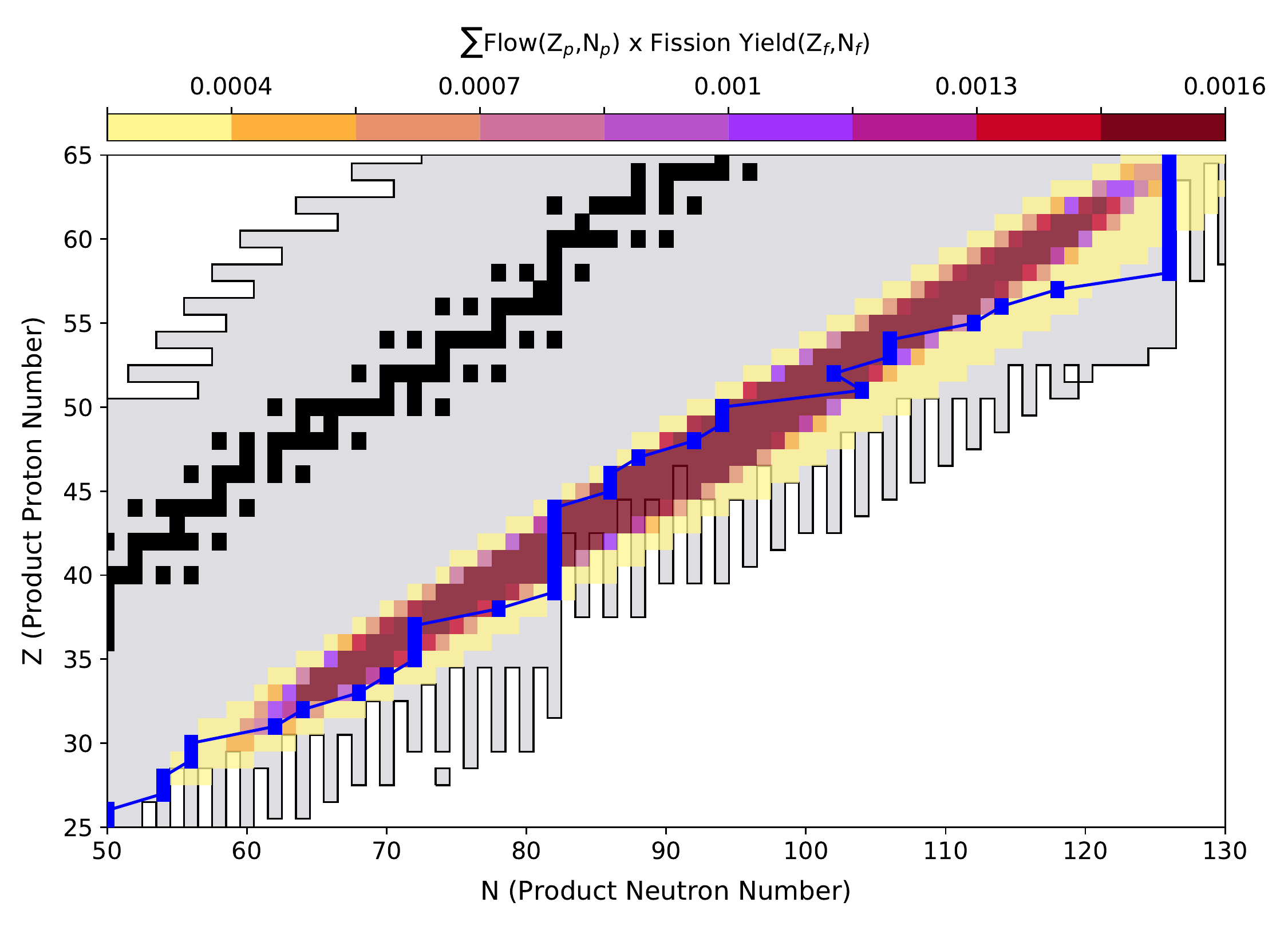}%
\hspace{0.001cm}
\includegraphics[width=8.8cm]{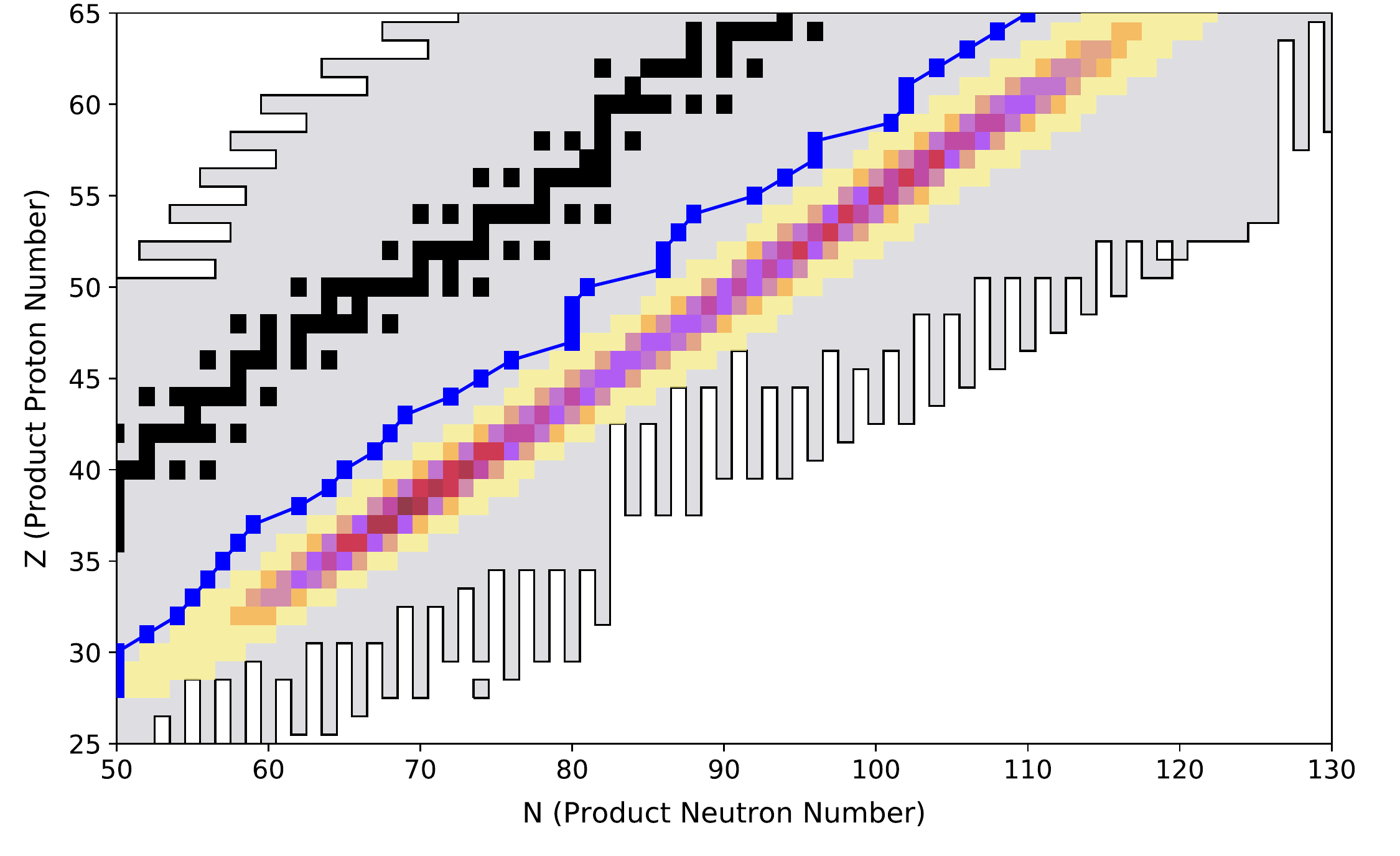}%
\end{center}
\caption{Fission flow of a parent nucleus ($Z_p$,$N_p$) cross referenced with the fission yield of the fissioning species to explicitly show deposition at early (top) and late (bottom) times. The fission deposition is compared to the location of the $r$-process path (blue) defined to be the most abundant species at a given element number.}
\label{fig:yielddeposition}
\end{figure}

We now consider where deposition occurs explicitly in order to demonstrate the mechanism by which FRLDM yields give robust contributions to lanthanide elements along with co-production of the light precious metals. In Figure~\ref{fig:yielddeposition} we consider the cold dynamical case at the early (1.035 seconds) and late (4.48 seconds) times denoted in Fig.~\ref{fig:yieldmetrics}. Here we take the fission flows of a given nucleus multiplied by the fission yield of the corresponding fissioning species to then sum the contributions to the possible products from all fissioning nuclei at a given time step. As can be seen in Fig.~\ref{fig:yielddeposition}, at early times deposition is spread broadly across $A\sim100-180$, with almost all of the deposition into the light precious metals to the left of the $r$-process path. With free neutrons still readily available at this early time, nuclei deposited here undergo neutron capture back to the path and the light precious metal region remains cleared out. In contrast, late-time deposition from mostly asymmetric yields introduces product nuclei that are found to the right of the $r$-process path. With free neutrons largely depleted, neutron-rich nuclei to the right of the path will primarily undergo $\beta$-decay, especially in cold conditions where photodissociation does not influence late-time dynamics, and thus these contributions remain in the light precious metal region. Therefore we find that it is the late-time deposition of light precious metals and lanthanides that most influences the final abundances in these regions of the pattern so that universality can be achieved without the need for many fission cycles. This is further supported by Fig.~\ref{fig:compconditions} where hot versus cold dynamical ejecta conditions see similar ratios among the light precious metals and the lanthanide elements although such conditions have very different late-time dynamics and fission cycling behavior. Although in such hot dynamical ejecta conditions the second $r$-process peak remains populated throughout the calculation, the light precious metals are still built up solely by late-time fission deposition that is achievable with a single fission cycle. The neutron-rich fission products to the right of the path in hot conditions are influenced by both photodissociation and $\beta$-decay, but nevertheless remain mostly in their late-time deposition location. 

\begin{figure*}
\begin{center}
\includegraphics[scale=0.515]{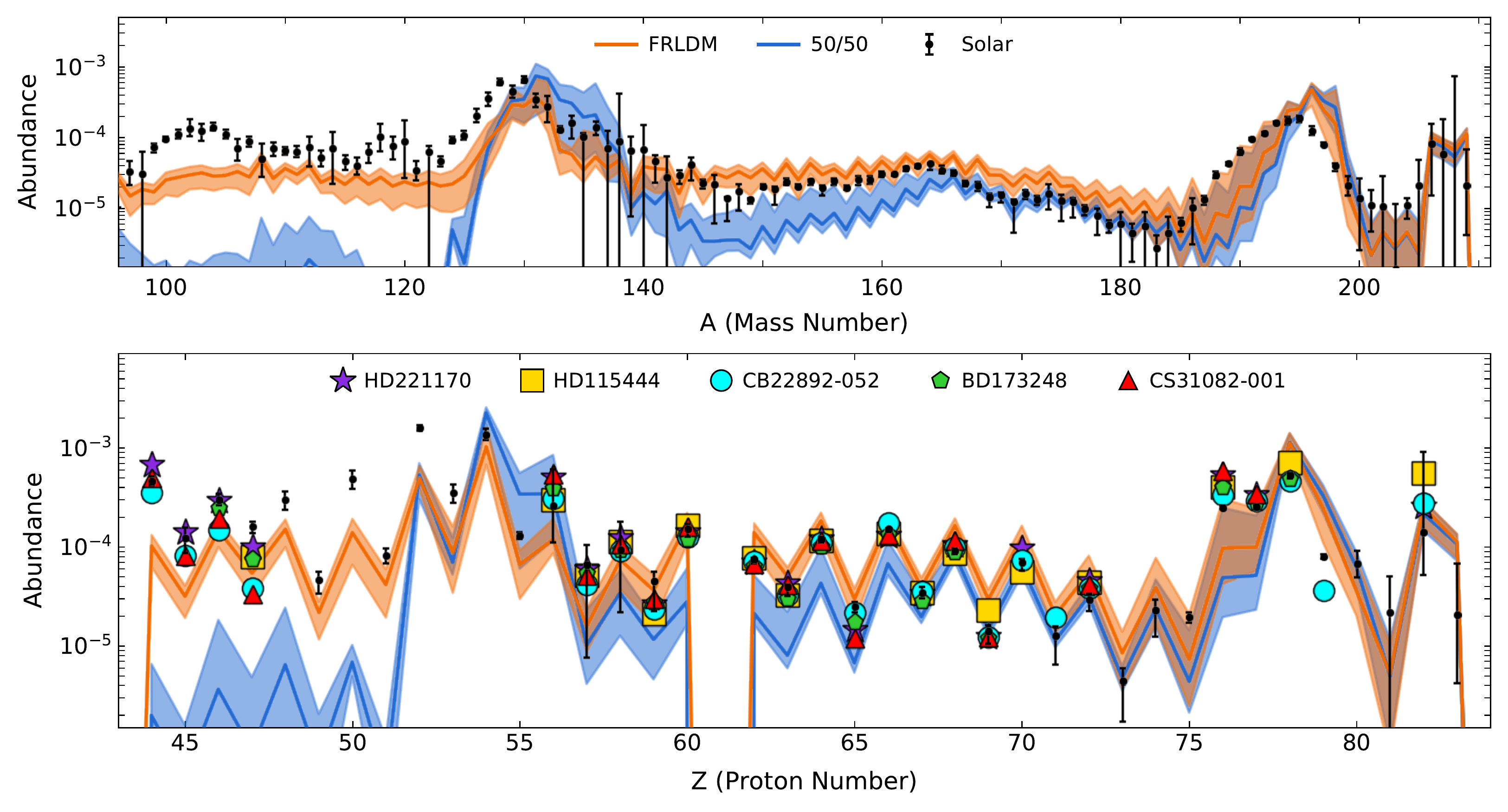}
\end{center}
\caption{The range (bands) and mass-weighted average (lines) of the $r$-process abundances for the dynamical ejecta simulation conditions of \cite{Rosswog} given 50/50 symmetric splits (blue) as compared to FRLDM fission yields (orange). Solar abundances and uncertainties \citep{goriely99,Arnould07} as well as abundances for the metal-poor, $r$-process rich stars considered in \cite{Sneden} are shown for comparison. Here the metal-poor star data is scaled by considering the ratio of their summed abundances between $Z=56$ and $Z=78$ to that found for solar data.}
\label{fig:yieldcomp}
\end{figure*}

The importance of the late-time fission deposition that takes place during the decay back to stability was previously emphasized in \cite{BDFrp,VasshGEF2019}. The exact influence of late-time fission depends strongly on an interplay between the fission yields and the fission rates. For instance although the fission rates determined using FRLDM barriers permit significant late-time contributions to the light precious metals given GEF2016 fission yields, applying HFB-14 barriers for the fission rates does not produce this behavior given this yield model (as can be seen in Figure 15 of \cite{VasshGEF2019}). However rates determined from HFB-14 barriers do permit some deposition into the light precious metals with other yield models such as SPY \citep{GorielyGEF} as well as with the the FRLDM yields explored here. In fact since FRLDM yields are wide for a broad range of neutron-rich nuclei, late-time deposition to the left of the second $r$-process peak occurs equally strongly when fission rates are determined from FRLDM versus HFB-14 barriers even though their differences imply a different set of late-time fissioning species to be of importance. We note that although with FRLDM fission barriers the empirical yield model GEF2016 can also produce nonnegligible contributions to the abundances of the light precious metals \citep{VasshGEF2019}, in this case fission deposition into this region is not as strong as is found with the FRLDM yields derived from macroscopic-microscopic theory.

We next comment on the potential for sites that host fission to produce a universal $r$ process by comparing simulation results directly with observational data for metal-poor, $r$-process enhanced stars. For this we show nucleosynthesis results given the mass-weighted average of thirty 1.2--1.4 $M_{\odot}$ neutron star merger dynamical ejecta trajectories from \cite{Rosswog} in Figure~\ref{fig:yieldcomp}, a subset of which were introduced in Fig.~\ref{fig:compconditions}. Here we compare results using FRLDM yields to the case where deposition is concentrated near $A\sim132$ when symmetric 50/50 splits are applied (these assume simple fission products having half the mass and atomic numbers of the fissioning species). Lanthanide abundances in the rare-earth region with FRLDM yields compare well with the observational data unlike the underproduction found with 50/50 splits. Additionally the FRLDM yield model sees over an order of magnitude more of an abundance of light precious metals than would be predicted assuming 50/50 splits. When comparing the elemental abundances of the light precious metals to the observational data, although ruthenium ($Z=44$) and rhodium ($Z=45$) are underproduced, we find elemental abundances for palladium ($Z=46$) and silver ($Z=47$) to be well reproduced by a yield model such as FRLDM that has deposition into this region at late times. We note that the isotopes of palladium found in the solar $r$-process residuals are at $A=105,106,108,$ and $110$ and isotopes of silver exist at $A=107$ and $109$. These are mass numbers at which fission deposition does not typically occur in substantial amounts given other yield models previously considered in $r$-process calculations \citep{Cowan}. Although \cite{Shibagaki} and \cite{GorielyGEF} demonstrated cases that saw broad fission deposition into both light precious metals and lanthanides, such deposition came at the expense of the second $r$-process peak with both cases having flat abundances in this region. However FRLDM results show that conditions that reach fissioning nuclei could simultaneously produce both a robust main $r$-process pattern along with light precious metals. This is of relevance when examining the possible link between fission and the universality of stellar and solar abundances since these stellar patterns must have a robust second peak in order to truly suggest a production mechanism similar to solar.

It should also be pointed out that with FRLDM yields depositing nuclei in a broad range across $A\sim100-180$, the details of the $r$-process abundance pattern are more sensitive to local features around $N=82$ such as shell effects or deformation, which can permit a local region to be extra stable relative to its neighboring nuclei. Therefore since there are presently many nuclear physics unknowns near the $N=82$ shell closure, a stronger shell closure than is predicted by FRDM2012 would produce a more pronounced second peak. Additionally, similar to the mechanism by which rare-earth peak formation in the lanthanide region could be achieved \citep{Reb97,Matt12,REMM1,REMM2,OrfordVassh2018}, should local masses or capture or decay rates have features that slow material down to the left of $N=82$, the raw material deposited here with FRLDM yields could pile up to further fill in the light precious metal peak elements such as ruthenium and rhodium.  

\begin{figure*}%
\begin{center}
\includegraphics[width=8.8cm]{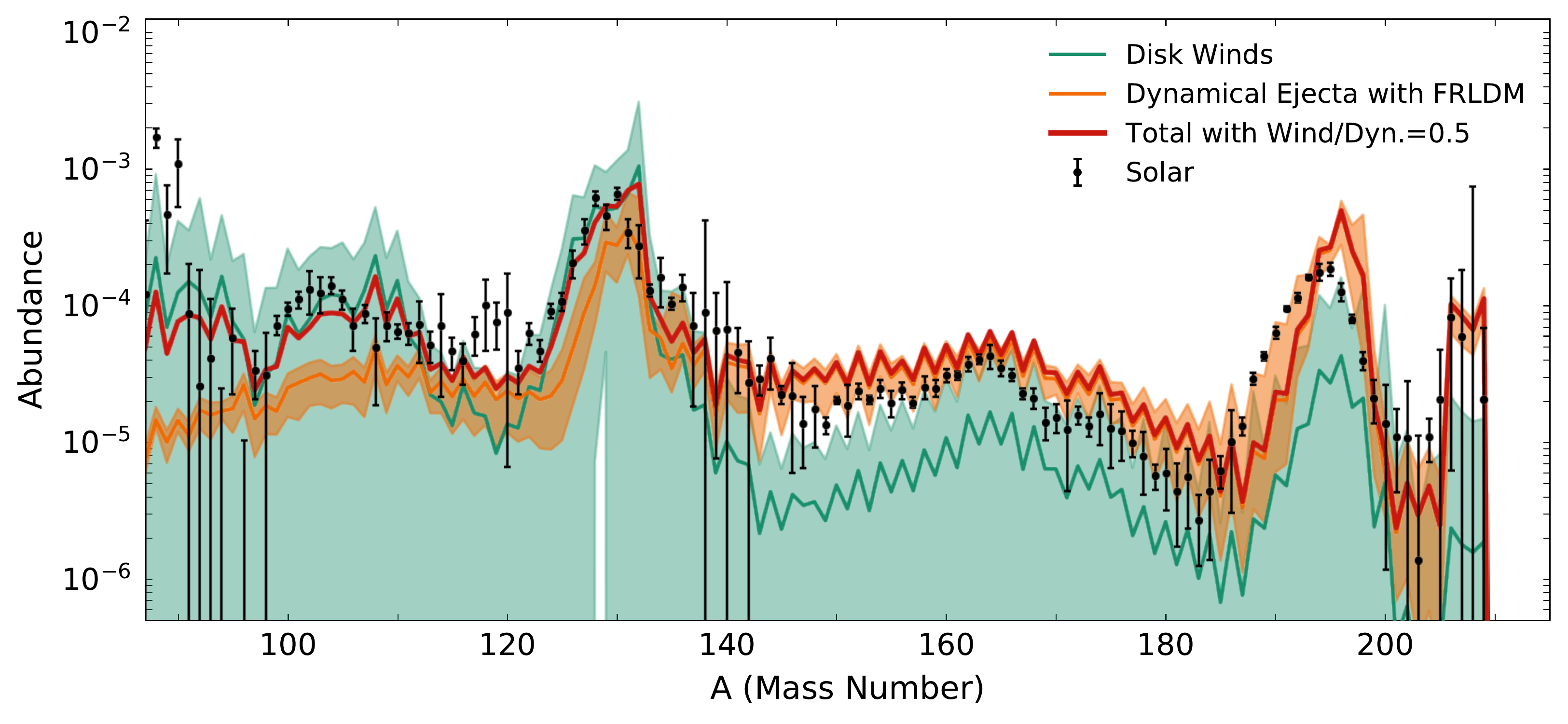} %
\includegraphics[width=8.8cm]{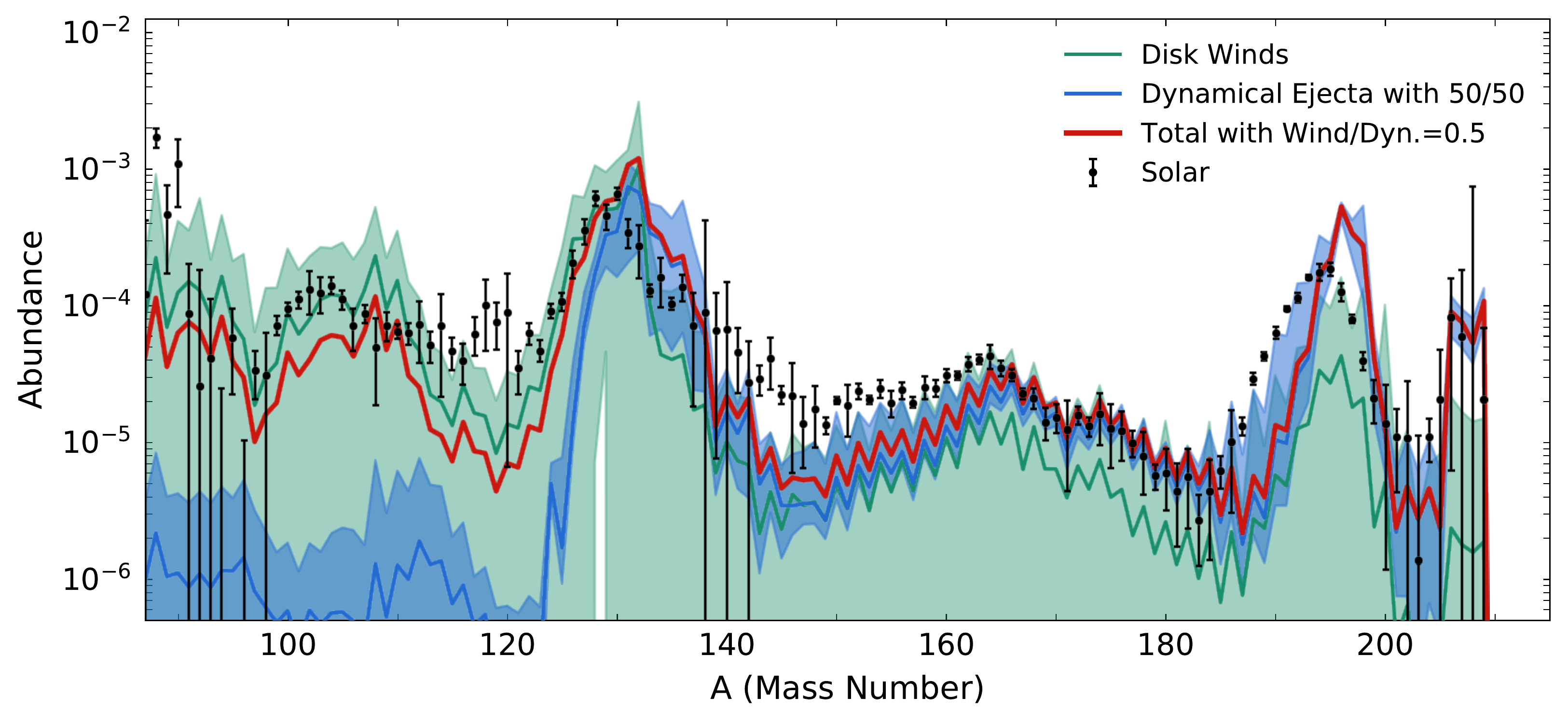}%
\hspace{0.01cm}
\includegraphics[width=8.8cm]{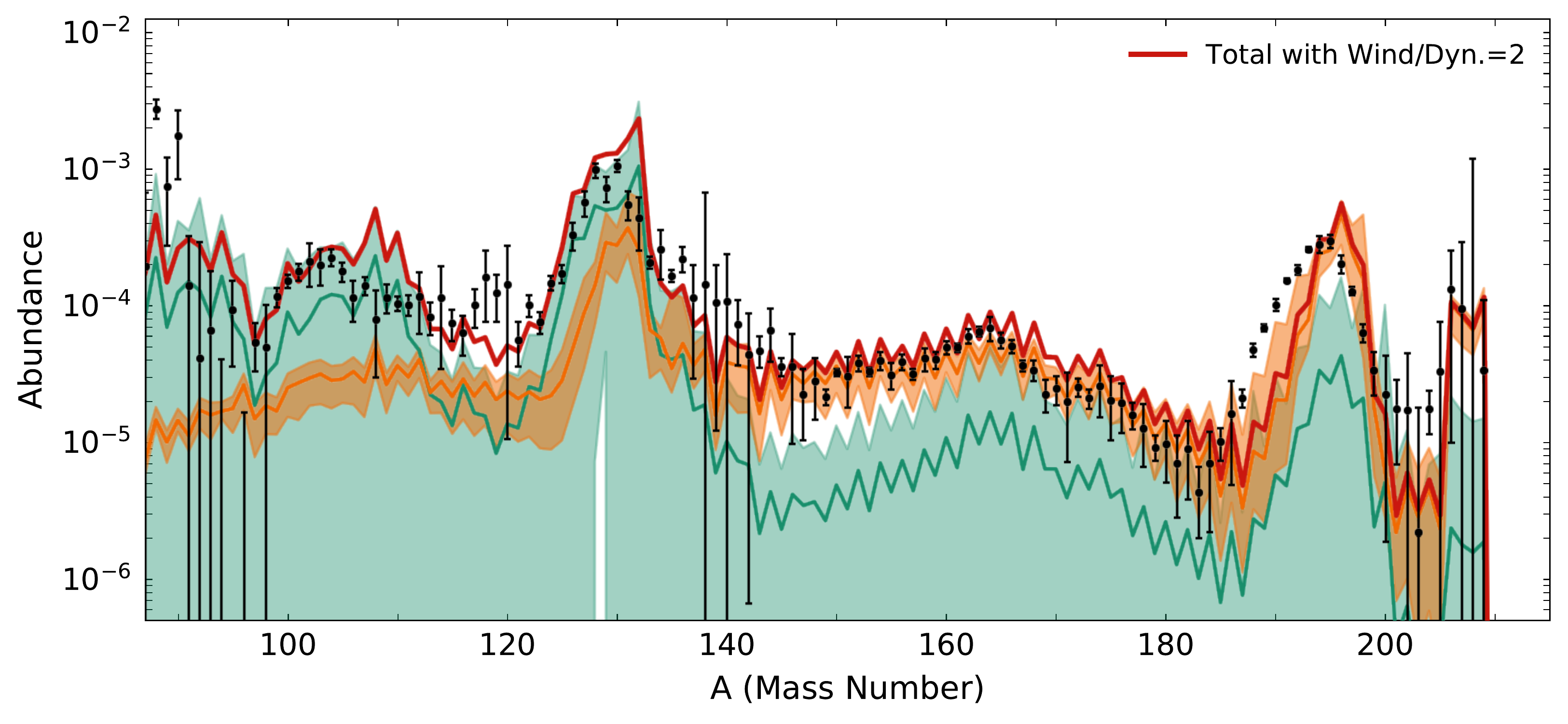}%
\includegraphics[width=8.8cm]{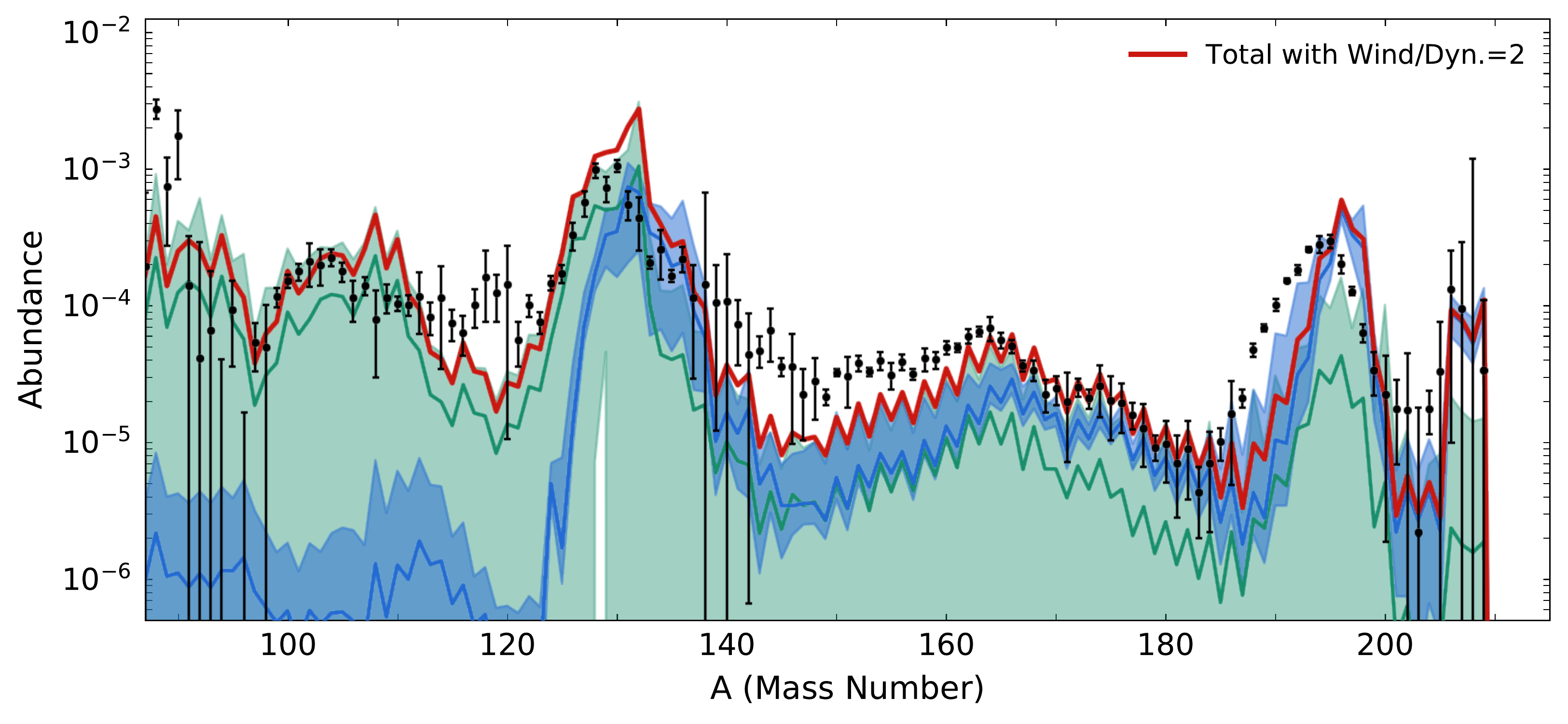}%
\end{center}
\caption{The total nucleosynthetic abundances from a merger event given a dynamical ejecta \citep{Rosswog} + accretion disk wind \citep{Just+15} scenario when the mass ratio of wind to dynamical ejecta is taken to be 0.5 (top panels) as compared to 2 (bottom panels). Orange bands (left panels) show results when FRLDM fission yields are applied as compared to 50/50 symmetric splits as blue bands (right panels). The solar data scaling is the same between 50/50 and FRLDM comparisons but distinct scaling is applied between the cases of differing dynamical to wind ejecta ratios.}
\label{fig:wind2dyn}
\end{figure*}

\section{Two-component merger model: dynamical ejecta and accretion disk winds}\label{sec:twocomp}

In addition to dynamical ejecta, the winds from the accretion disk that later forms around the central remnant will also contribute to the mass ejection from merger events. Such accretion disk winds can contribute anywhere from a weak to strong $r$ process depending on the conditions \citep{Just+15,Perego+14}. Although wide ranges on the ratio of wind ejecta to dynamical ejecta in such events have been predicted \citep{CoteGW170817}, it remains likely that both components participate in the nucleosynthetic outcome.

In Sec.~\ref{sec:rpimpact} we considered the case where dynamical ejecta have the very low $Y_e$ conditions that are favorable for fission and found that late-time fission deposition can significantly contribute to the abundances of the light precious metals, specifically palladium and silver. Here we investigate whether the solar abundance pattern can accommodate the production of such nuclei via fission when it occurs alongside weak $r$-process contributions. We use an accretion disk wind for weak $r$-process abundances in order to consider a two-component merger model where both dynamical and wind ejecta contribute, however we note that the lighter heavy element abundances in the solar pattern could also reflect contributions from a LEPP as well as more processed higher $Y_e$ dynamical ejecta (the later possibility is explored in detail in Section~\ref{sec:radicedyn}). For the accretion disk wind, we use 2092 viscously driven wind tracers from the M3A8m3a2 simulation of \cite{Just+15} for which fission does not influence the final abundances. This wind case populates the light precious metals, such as silver, along with lanthanides, such as europium, via mass-weighted combinations of simulation tracers that separately contribute to these regions by undergoing either a weak or main $r$ process.

We compare results given FRLDM and 50/50 fission yields for such a two-component merger model in Figure~\ref{fig:wind2dyn}. When dynamical ejecta is taken to be twice as plentiful as wind ejecta, and FRLDM fission yields are considered, the abundances of light precious metals and heavier $r$-process nuclei are easily accommodated. In contrast, when 50/50 splits are instead considered along with a ratio dominated by dynamical ejecta, the light precious metal region is greatly underproduced. Results with precisely symmetric 50/50 fission splits therefore require more wind ejecta to explain the production of the light precious metals, demonstrated by the figure panel where the ratio of wind to dynamical ejecta is taken to be two. However, for this ejecta ratio results using the FRLDM yield model still well reproduce the full pattern and help to fill in the troughs of absent material on the left and right of the second $r$-process peak.

We find that a fission yield model such as FRLDM that predicts broad fission deposition around the second $r$-process peak permits a reproduction of the full $r$-process pattern even when contributions from fission products are accompanied by various amounts of weak $r$-process ejecta, as evidenced by the robustness of the pattern for the total ejecta in the presence of variable ratios of wind to dynamical ejecta shown in Fig.~\ref{fig:wind2dyn}. By depositing into both the light precious metals and lanthanides, the FRLDM fission yield model decreases the sensitivity of merger event outcomes to the exact ratio of wind to dynamical ejecta. Late-time fission contributions in the light heavy element region therefore help to stabilize the abundances against potentially naturally occurring variations in merger ejecta conditions. Thus although it is possible for solar and stellar patterns to be accommodated without a contribution from fission to the light precious metals via weak $r$-process and main $r$-process ejecta produced in exactly the right ratios, we find fission to be an alternative or complementary mechanism to achieve consistency in abundances.

\section{Dynamical ejecta with weak and strong \lowercase{$r$}-process contributions}\label{sec:radicedyn}
Simulations of neutron star merger dynamical ejecta often find very neutron-rich conditions favorable for a fission cycling $r$ process, although the exact amount of such ejected matter remains debatable. Some simulations show a broad range in dynamical ejecta conditions with very low $Y_e$ ejecta only constituting a fraction of the total ejecta, while other simulations suggest such a low $Y_e$ component to dominate dynamical ejecta \citep{Radice18,Bovard,Wanajo+14}. Here we make use of astrophysical conditions from a recent simulation of merger dynamical ejecta given the SFHo equation of state for a 1.2--1.4 $M_{\odot}$ neutron star merger that takes into account the influence of weak interactions via M0 neutrino transport (which goes beyond leakage schemes when treating neutrino absorption) \citep{Radice18}. In contrast to the dynamical ejecta conditions of \cite{Rosswog} considered in previous sections, which all represented low entropy, low $Y_e$ conditions in which fission robustly participates, the dynamical ejecta model considered here contains a broad range of astrophysical conditions such that weak $r$-process contributions occur alongside strong $r$-process contributions. We investigate how fission contributes to the nucleosynthetic outcome in such a case.

\begin{figure}
\begin{center}
\includegraphics[scale=0.67]{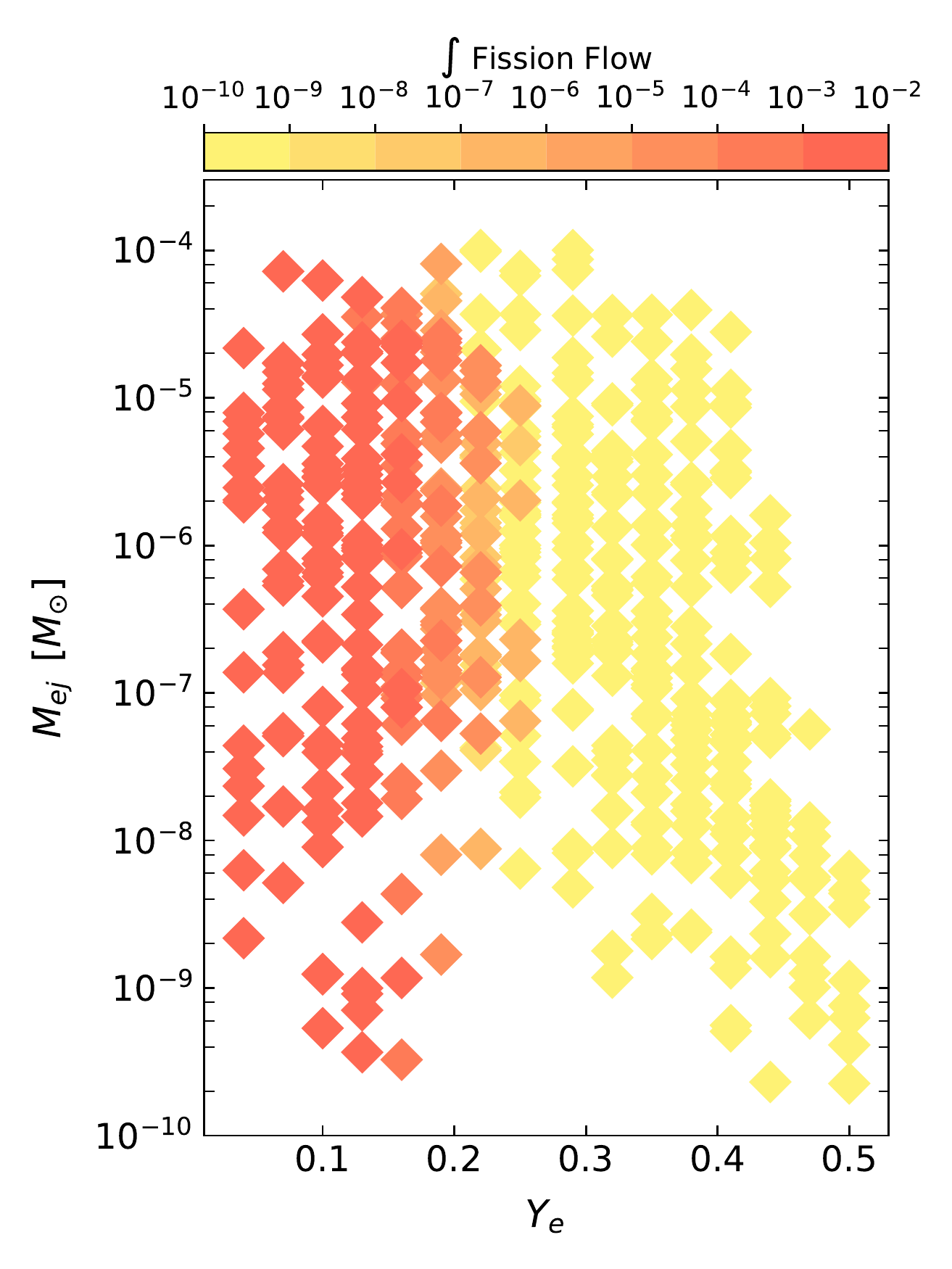}
\end{center}
\caption{The amount of ejecta mass, $M_{ej}$, and initial neutron richness, $Y_e$, for the dynamical ejecta of a 1.2--1.4 $M_{\odot}$ neutron star merger simulation \citep{Radice18} that uses the SFHo equation of state and M0 neutrino transport. Each case is color coded by the time integrated total fission flow (rate times abundance) found for each of these trajectories from nucleosynthesis calculations by summing the contributions from both $\beta$-delayed and neutron-induced fission.}
\label{fig:radiceyemejfisscol}
\end{figure}

The participation of fission in all trajectories from this simulation is represented in Figure~\ref{fig:radiceyemejfisscol} via color-coding each case by the time integrated neutron-induced and $\beta$-delayed fission flow (where flow is rate multiplied by abundance). Fissioning nuclei are robustly accessed in trajectory conditions with $Y_e<0.1$, however it is not solely the neutron richness that determines the reach of the $r$ process since the initial entropy and dynamical timescale determining the temperature and density evolution influence the outcome. As can be seen in Fig.~\ref{fig:radiceyemejfisscol}, depending on the entropy and timescale, conditions with $0.1<Y_e<0.25$ can also access fissioning nuclei. We emphasize that our detailed considerations of individual trajectories here are to serve as a means to understand the possible influence of fission on the total mass-weighted ejecta. It is evident that although weak interactions produce a fraction of ejecta that will only undergo a weak $r$ process and not access fissioning species, a significant component of the ejecta that hosts fission remains.

\begin{figure}%
\begin{center}
\includegraphics[width=8.725cm]{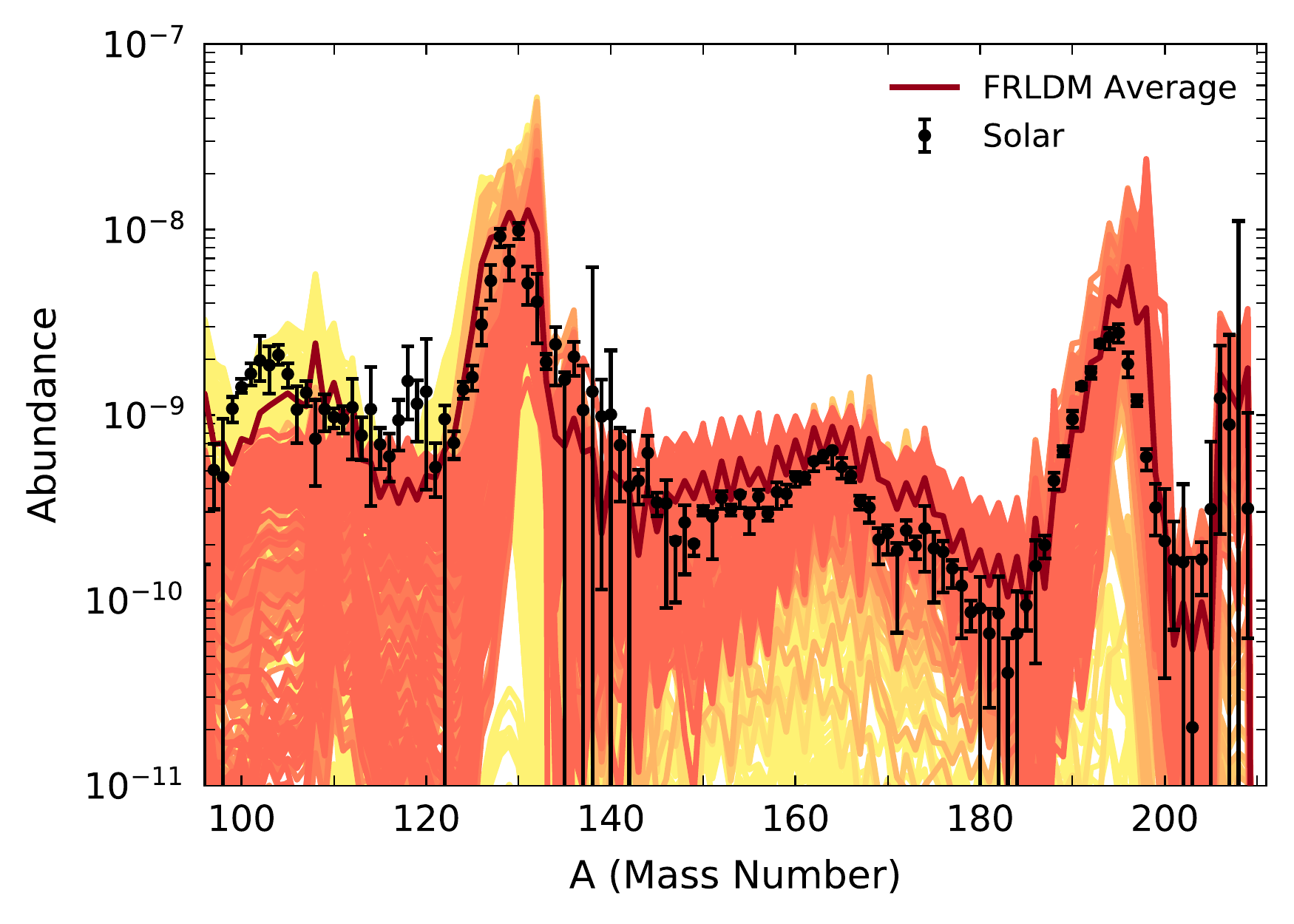}%
\hspace{0.001cm}
\includegraphics[width=8.725cm]{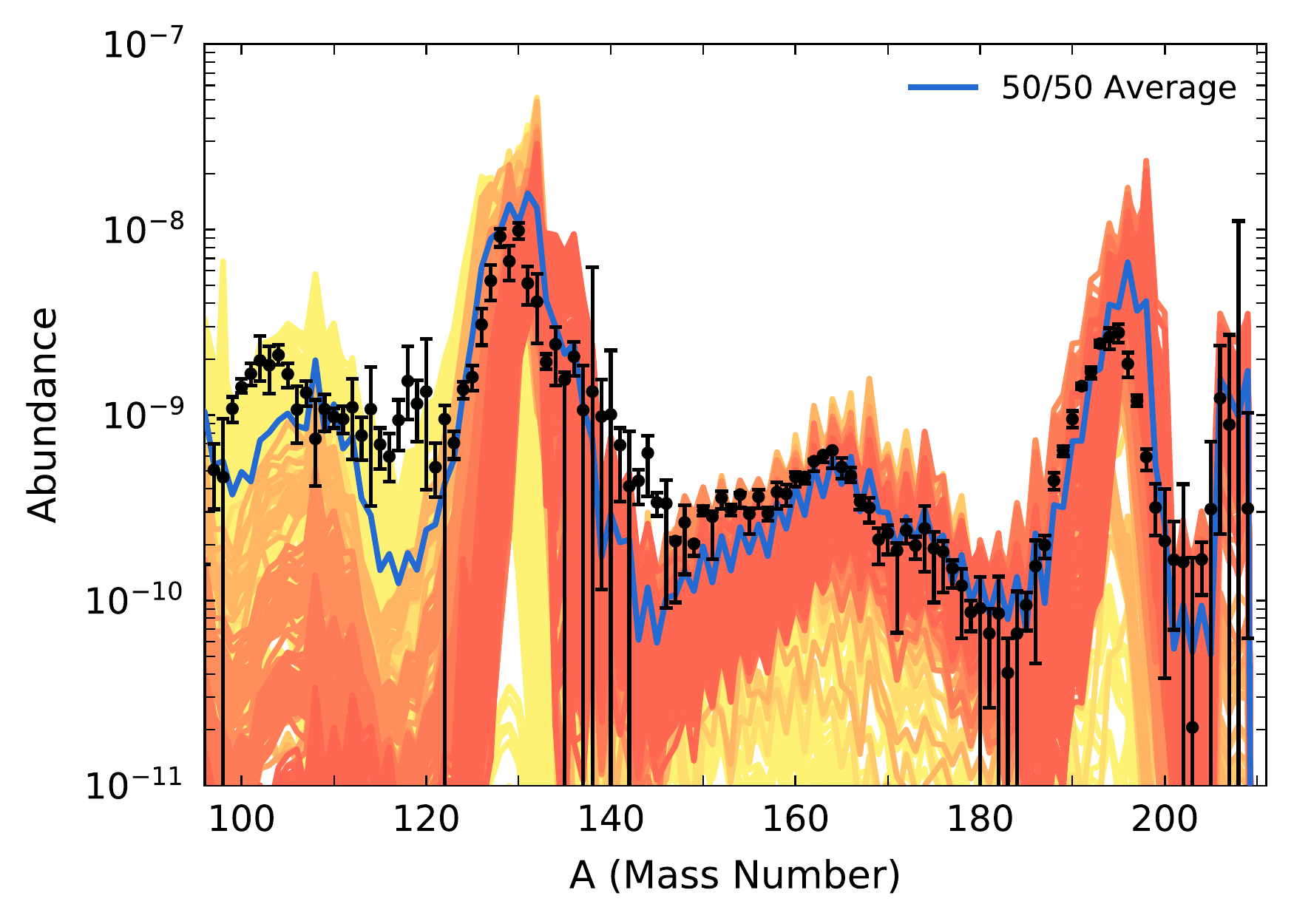}%
\end{center}
\caption{The final $r$-process abundances for all simulation trajectories shown in Figure~\ref{fig:radiceyemejfisscol} color coded by their time integrated fission flow (same color scale as in Fig.~\ref{fig:radiceyemejfisscol}) for the cases of FRLDM (top) and 50/50 (bottom) fission yields. The total mass-weighted abundance is also shown by the red and blue bold lines for FRLDM and 50/50, respectively.}
\label{fig:radiceabAfisscol}
\end{figure}

The $r$-process abundances for each of the trajectories seen in Figure~\ref{fig:radiceyemejfisscol} are shown in Figure~\ref{fig:radiceabAfisscol} along with the mass-weighted average given FRLDM and 50/50 fission splits. Dynamical ejecta conditions that do not reach fissioning nuclei can populate the light precious metal region and in the case of 50/50 splits are the only means by which dynamical ejecta form the lighter heavy elements. In contrast, when wide fission yields that deposit to the left of the second peak such as FRLDM are considered, fission products are an important contributor to the abundances of the light precious metals.

\section{Hints of co-production from light and heavy \lowercase{$r$}-process elements in \lowercase{r}-I and \lowercase{r}-II stars}\label{sec:rstars}

Elemental ratios are often explored with the hope that a dependence on a particular nucleosynthetic process or site will emerge and shed light on the evolution of elements in the Galaxy. For instance low-metallicity $s$-process rich stars show much larger values for their barium to strontium, [Ba/Sr], ratio then very $r$-process rich stars \citep{Sneden}. Another widely explored ratio is that of europium over iron, [Eu/Fe], which has been used to consider the possible contributions mergers can make to Galactic yields since observation suggests that iron production from supernova must drive this ratio down at later times \citep{CoteEichler}. Additionally the downward trend in strontium over europium, [Sr/Eu], as a function of [Eu/Fe] observed in metal poor stars has been argued to suggest that light heavy element enrichment just after the first $r$-process peak can be high even when main $r$-process enrichment is low implying a more frequent event to be mostly responsible for the production of strontium \citep{MontesApJ07}.

Here we further consider the scenarios outlined in Sections~\ref{sec:twocomp} and \ref{sec:radicedyn} by investigating elemental ratios for a lanthanide element just beyond the second peak, lanthanum (La, $Z=57$), which is often considered representative of the robustness or universality argument \citep{Sneden} as compared to elements found in the light precious metal peak of $100<A<111$ produced mostly by isotopes of ruthenium (Ru), but also rhodium (Rh), palladium (Pd), and silver (Ag). Specifically here we focus on the heaviest of such elements, palladium and silver, in order to consider the implications when trends from the observational data for metal-poor stars are compared to nucleosynthetic predictions using FRLDM fission yields that find nonnegligible contributions to the left of the second $r$-process peak from fission deposition. We therefore compare only to the observational trends in $r$-process enhanced r-I and r-II stars in order to probe the cases that have synthesized main $r$-process elements. Here we use the europium abundance, log eps(Eu), as a proxy for the $r$-process enrichment of the star and adopt the europium criterion to classify r-I and r-II stars as well as the definition of metal-poor from \cite{AbohalimaFrebel}. Note that although we find fission deposition to be distributed past the light precious metal region and leading into the second peak and beyond, little to no observational data for lighter heavy elements beyond silver such as cadmium presently exist.

\begin{figure*}%
\begin{center} 
\includegraphics[width=8.8cm]{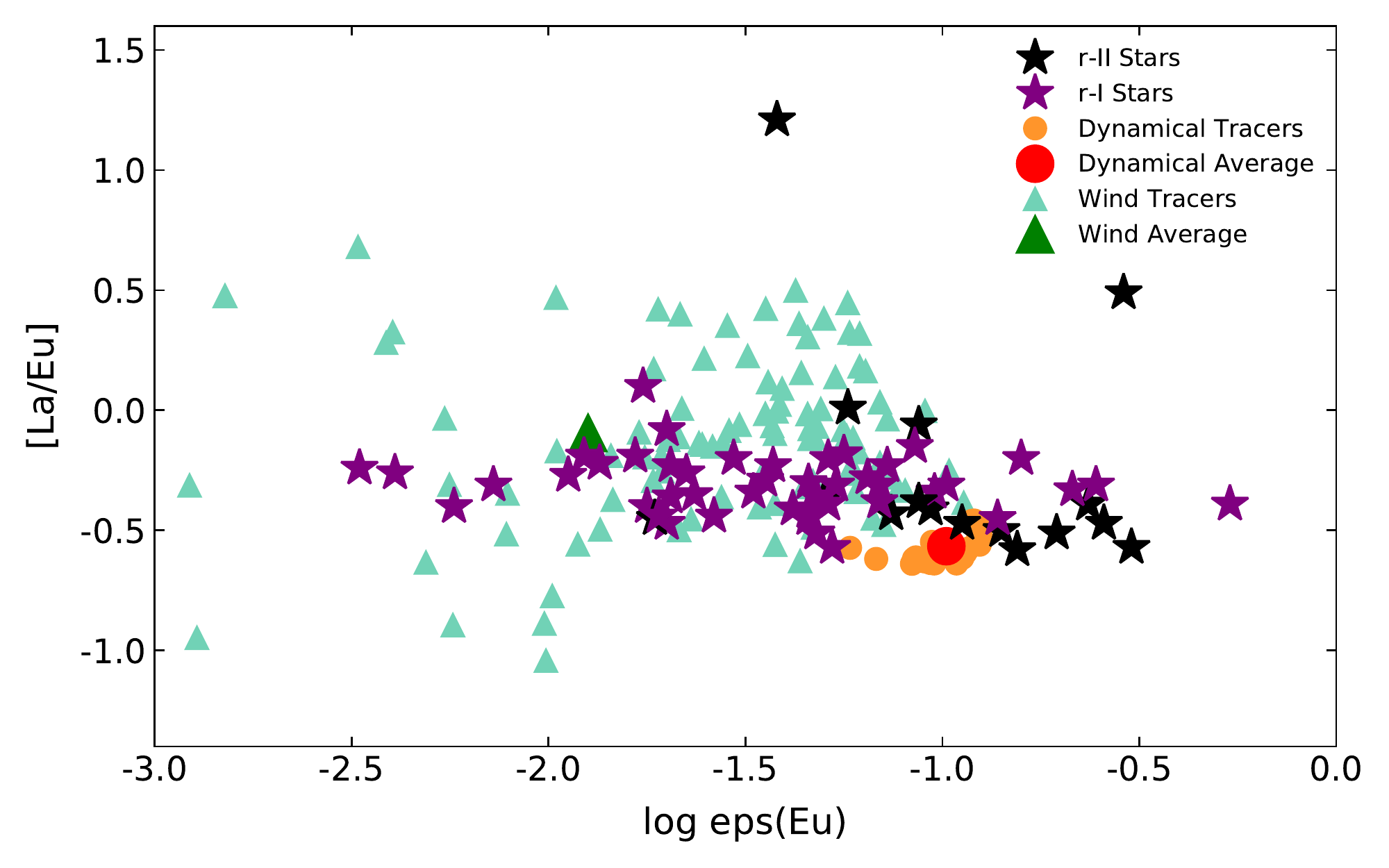} %
\includegraphics[width=8.8cm]{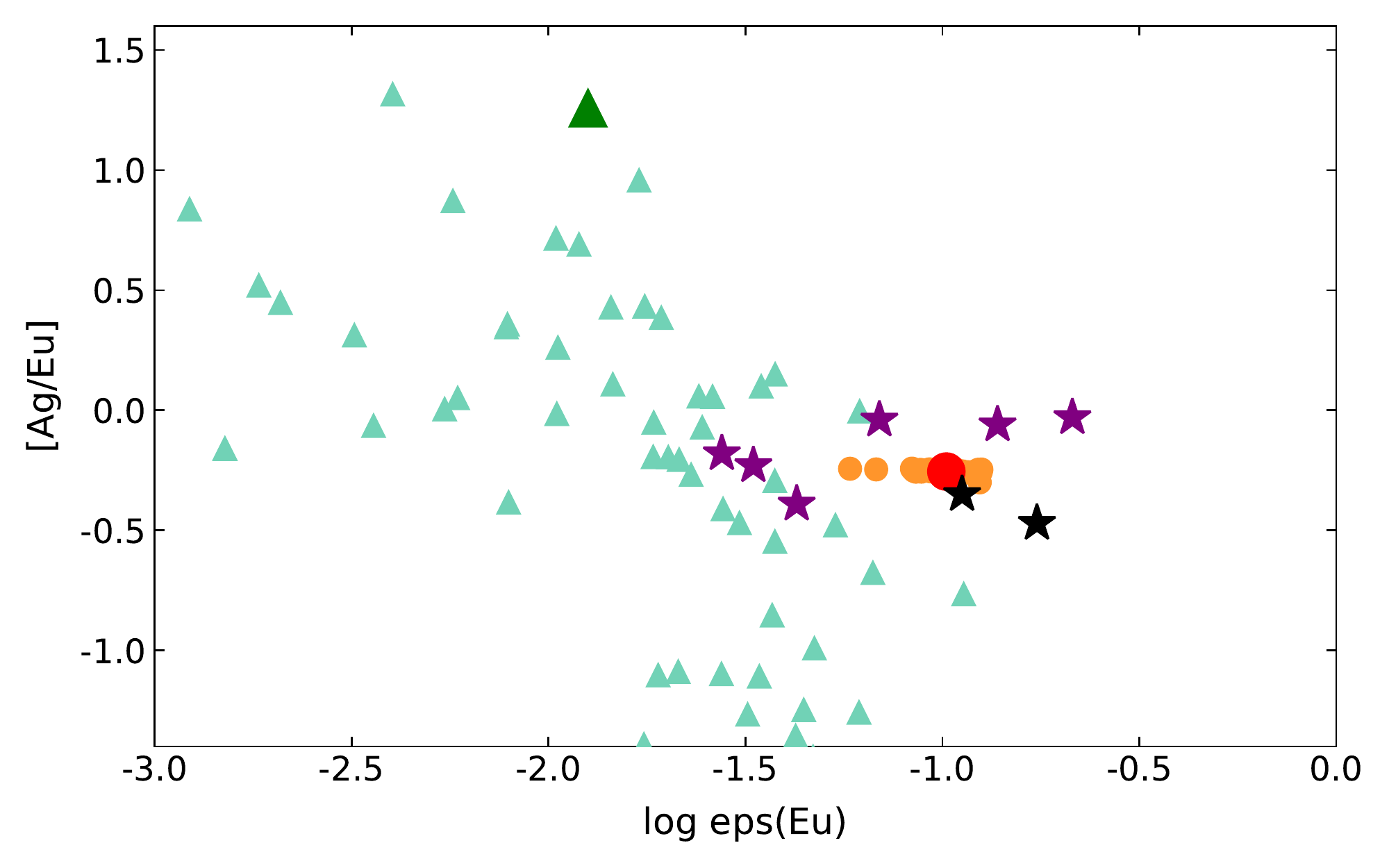}%
\hspace{0.5cm}
\includegraphics[width=8.8cm]{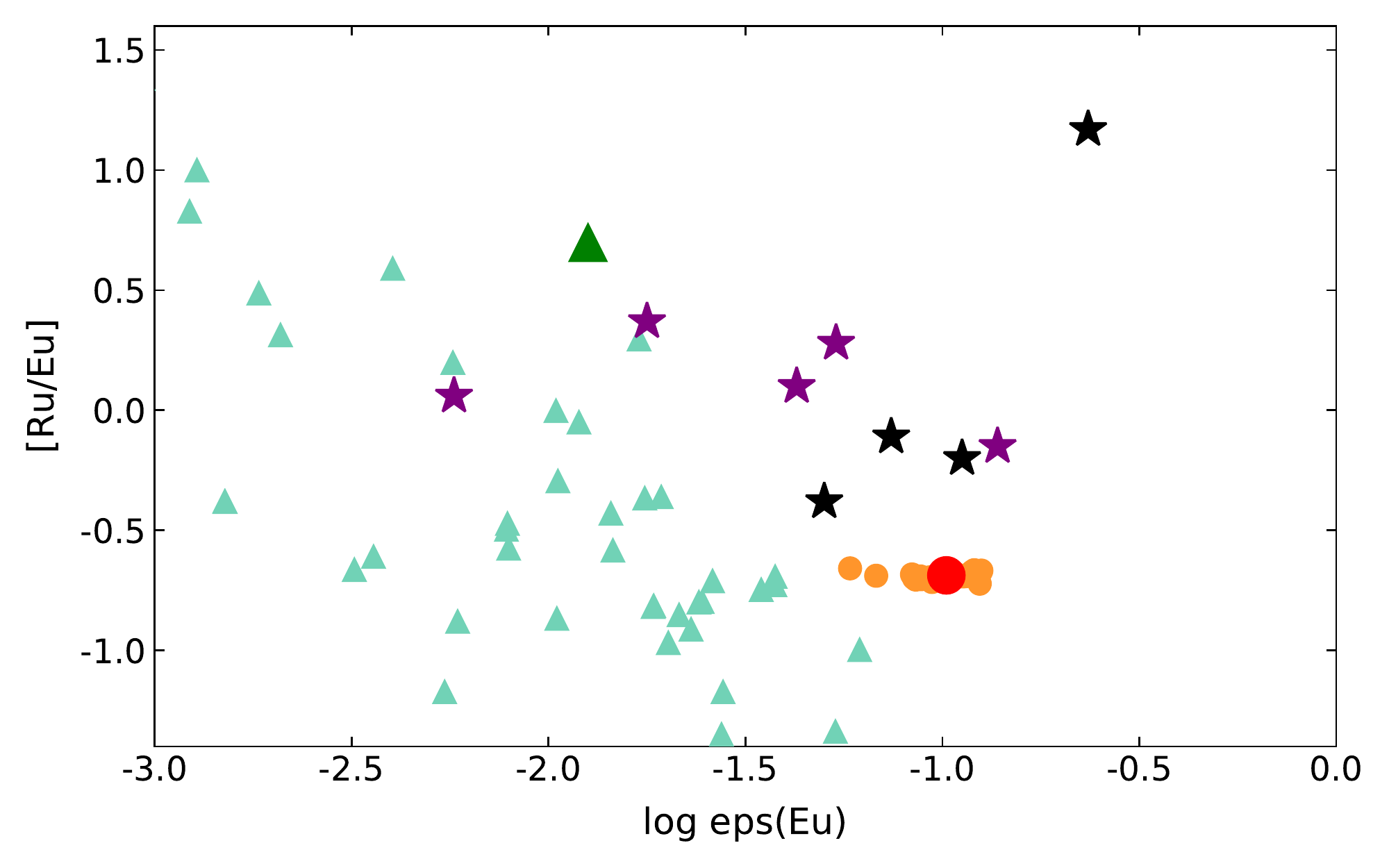} %
\includegraphics[width=8.8cm]{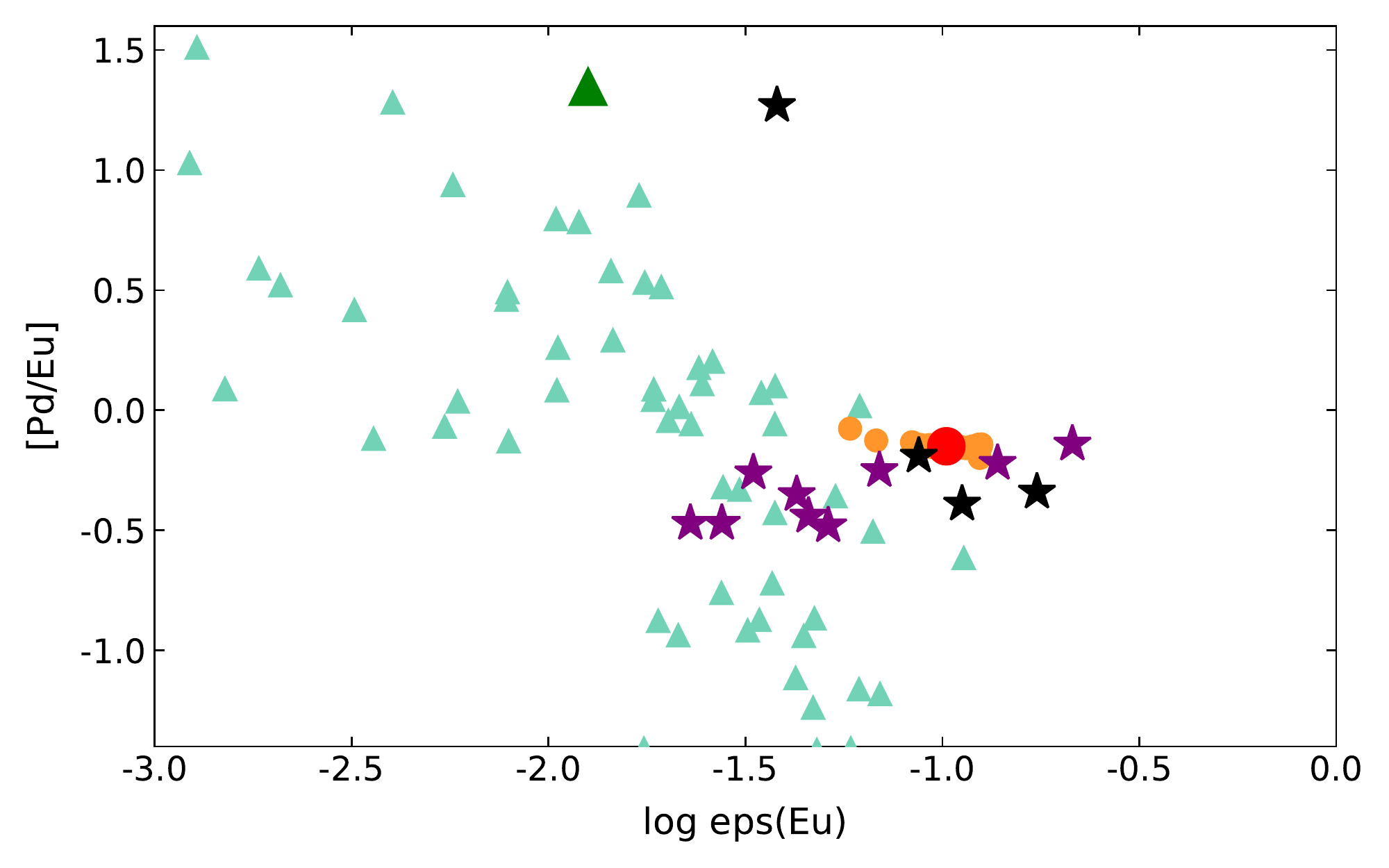} %
\end{center}
\caption{The elemental ratio for the lanthanide element lanthanum over another lanthanide element europium (top left) as compared to elemental ratios for the light precious metals silver (top right), palladium (bottom right), and ruthenium (bottom left) over europium. Triangles show results from an accretion disk wind simulation mass-weighted average (dark green) as well as a subset of individual tracers (light green). Circles show results from the merger dynamical ejecta tracers of \cite{Rosswog} (orange) as well as the mass-weighted average (red) when FRLDM fission yields are applied. The observational data for metal-poor r-I (purple) and r-II (black) stars are taken from JINAbase \citep{AbohalimaFrebel}. Here a uniform scaling is applied to the simulation tracers and the metal-poor star data reflect their observed enrichment.}
\label{fig:rstars}
\end{figure*}

We first make use of the two-component scenario discussed in Section~\ref{sec:twocomp} which considered an accretion disk wind along with low $Y_e$ dynamical ejecta conditions in order to highlight the stark contrast in elemental ratios for cases that primarily produce a weak $r$ process as compared to conditions that robustly access fissioning nuclei. The [La/Eu] ratio shown in Figure~\ref{fig:rstars} demonstrates the universality argument frequently discussed for elements with $Z\ge56$. Here the flat trend shows that metal-poor, r-rich stars exhibit approximately the same ratio for lanthanum to europium regardless of the exact enrichment of the star, suggesting that the same type of event has polluted the environments in which these stars formed and that such an event produces a robust $r$-process pattern. In contrast, for Ru, the lightest of the light precious metals considered in Fig.~\ref{fig:rstars}, the order of magnitude variance seen in the [Ru/Eu] ratio for r-I and r-II stars suggests Ru production occurs independently of Eu production, which could arise from the presence of multiple sources, contributions from a LEPP, or a greater sensitivity to the exact merger conditions such as progenitor mass as compared to elements populated by a main $r$ process such the lanthanides.

\begin{figure*}
\begin{center}
\includegraphics[scale=0.6]{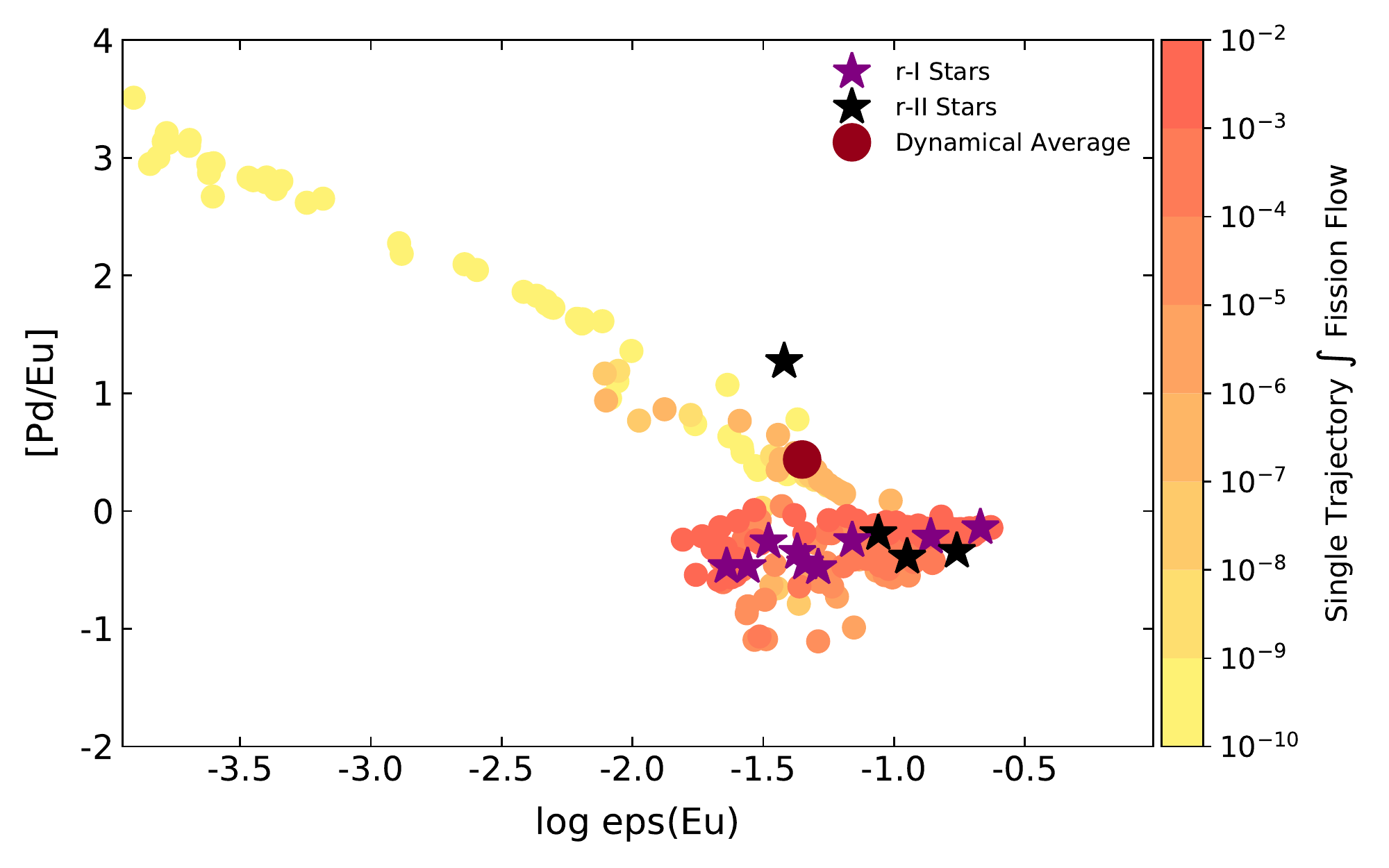}
\end{center}
\caption{Same as the bottom right panel of Fig.~\ref{fig:rstars} but showing instead nucleosynthesis calculation results using dynamical ejecta trajectories from \cite{Radice18} that display a broad range of astrophysical conditions. Ratios obtained for individual trajectories are color coded by their total time integrated fission flow (rate times abundance). Due to late-time deposition, the trajectories with a strong influence from fission are those that readily accommodate the trend seen in r-I and r-II stars.}
\label{fig:radiceratiosfisscol}
\end{figure*}

When the heaviest elements in the light precious metal peak, Pd and Ag, are considered however a flat trend emerges for $r$-process enhanced stars similar to the behavior for [La/Eu]. This suggests such elements and Eu could be correlated and therefore co-produced. To compare to predictions from our nucleosynthesis calculations, we first consider the sample of wind simulation tracers with masses between (1--2) $\times\,10^{-5}$ solar masses (318 tracers in total) which is the most commonly populated mass ejection range for this simulation. The [Pd/Eu] and [Ag/Eu] ratios predicted by the accretion disk wind tracers here show a large spread and an overall downward trend. Since the conditions associated with each simulation tracer could be produced in different mass-weighted distributions given naturally occurring variations such as progenitor masses, timescale of formation of the remnant hypermassive neutron star or black hole, and black hole spin, astrophysical scenarios such as the accretion disk wind case studied here that rely entirely on combinations of ejected mass to populate both the light precious metals and lanthanides could see significant variability in their production of Pd and Ag relative to Eu, which is in tension with the consistency seen in the observational data. Additionally, the mass-weighted average of all wind simulation tracers lies well above the observational values demonstrating that such conditions frequently overproduce the light precious metals, such as Pd and Ag, relative to the lanthanides such as Eu. 

In contrast to the accretion disk wind simulation results, the values for [Pd/Eu] and [Ag/Eu] we find with the low $Y_e$ dynamical ejecta from \cite{Rosswog} that see co-production of such elements via fission deposition are remarkably consistent with observational ratios. We note that the consistency with the observational data also extends to comparisons with the solar ratios for [Pd/Eu] and [Ag/Eu] (as evidenced in Fig.~\ref{fig:yieldcomp}).  Although for the case of Pd there exists a single strong outlier, star SDSSJ103649.93+121219.8, we emphasize it is the overall average trend in the stellar data which is relevant to illuminate elemental dependencies on production sites. Here we show not only the ratio determined by a mass-weighted average of ejecta but also the values given by the individual tracers to emphasize that the trend holds for the full range of conditions present in this low $Y_e$ ejecta. Although the spread in Eu enrichment seen in metal-poor, r-rich stars is likely largely due to inhomogeneous mixing within the environment where these stars form \citep{JiMixRecII}, the simulation tracer distribution shown in Fig.~\ref{fig:rstars} for dynamical ejecta mirrors the trend in the observational data, providing another possible path to accommodate the spread observed in such stars since different merger events would likely see a variance in the astrophysical conditions. 

Lastly we emphasize that the dynamical ejecta case considered in Figure~\ref{fig:rstars} is simply an example of conditions with which the nucleosynthetic outcome strongly depends on fission while the accretion disk wind demonstrates a case in which fission does not influence final abundances. Although this is generally consistent with the current state of accretion disk wind and dynamical ejecta simulations, it still remains possible that cases in which wind ejecta outcomes are strongly influenced by fission exist in nature (recall the wind case with fission demonstrated in Fig.~\ref{fig:compconditions}). Therefore our results are not meant to suggest that merger dynamical ejecta alone are the source for Pd and Ag in r-I and r-II stars. Rather, to accommodate the co-production behavior suggested by r-I and r-II stars all that is needed is for Pd, Ag, and lanthanide abundances to be largely determined by late-time asymmetric fission yields instead of being entirely built up by separate combinations of conditions that could naturally vary. This point is further supported in Fig.~\ref{fig:radiceratiosfisscol} which considers the [Pd/Eu] elemental ratio for the dynamical ejecta simulation discussed in Section~\ref{sec:radicedyn}. Here it is evident that the conditions that most heavily access fissioning nuclei are those most consistent with the ratio observed in $r$-process enhanced, metal-poor stars. Trajectories which instead do not reach fissioning nuclei tend to overproduce Pd relative to Eu. Since this simulation case predicts a slightly higher contribution from conditions that do not reach fissioning species, the mass-weighted average value predicted for the [Pd/Eu] ratio is high relative to observation. Therefore this could indicate that dynamical ejecta scenarios should have a higher mass weighting for conditions with late-time fission deposition in order to accommodate the observed [Pd/Eu] flat trend indicative of co-production. We emphasize that the individual trajectory ratios shown in Figs.~\ref{fig:rstars} and \ref{fig:radiceratiosfisscol} are not meant to suggest that solely a fraction of the ejecta is responsible for the enrichment of these stars. Rather our results suggest that a higher mass weighting of ejecta that is affected by fission may be a means to explain the stellar trends.

\section{Conclusions}\label{sec:conclude}

Fission cycling as an explanation for the so-called universality or robustness of abundance patterns seen in $r$-process enhanced, metal-poor stars as compared to our Sun remains an intriguing prospect. We have demonstrated that the FRLDM fission yields \citep{Mumpower+19} deposit product nuclei in a wide range from the light precious metal region leading into the lanthanides from $100\lesssim A\lesssim 175$ and $44\lesssim Z\lesssim 71$. This wide range of deposition can help stabilize the pattern against fluctuations in the specifics of the astrophysical scenario, such as the exact ratio of wind to dynamical ejecta. Additionally, although the shape and height of the second $r$-process peak are influenced by the exact conditions present in the ejecta, the abundances to the left and right of the second $r$-process peak, where observational data are suggestive of universality, are fairly consistent in all fission dependent dynamical ejecta scenarios, making a case for a possible connection between universality and fission.

We have demonstrated that the flat trends in the elemental ratios suggestive of co-production are not solely seen for elements with $Z\ge56$ since observational data trends for palladium and silver from metal-poor, r-rich r-I and r-II stars show similar behavior. Therefore it is possible that the universality argument could be extendable down to the heaviest of the light precious metals, palladium and silver, so that such abundances can be explained in the case of $r$-process enhanced stars without invoking a LEPP, although a LEPP or other weak $r$-process sources seem to be required for elements lighter than this, such as ruthenium and rhodium, since the stellar variances here are not consistent with universality. We have shown that late-time deposition of asymmetric fission yields, such as is seen with the new wide fission yields of the FRLDM model, provides a way to explain the stability in Pd, Ag, and lanthanide elemental ratios from star to star due to significant abundance contributions to many isotopes around the second $r$-process peak. However, since the trends seen in the observational data are based on a small sample of stars, further observations of Pd and Ag are needed to gather statistics and confirm the suggested co-production behavior. Therefore, investigations such as those presented in this work will benefit greatly from observational efforts such as those being undertaken by the R-Process Alliance (RPA) \citep{RPANorth,RPASouth} to locate and analyze larger samples of $r$-process enhanced stars. Additionally, if observations could push beyond silver leading into the second $r$-process peak, e.g. \cite{RoedererTellurium}, co-production could be further tested and help to constrain yield models if indeed a fission cycling $r$-process site is the event responsible for the heavy element content of such metal-poor, r-rich stars. 

Our findings demonstrate the value of theoretical efforts to understand the fission properties of neutron-rich nuclei. There is still much progress to be made with the nuclear data in the future. For instance, spontaneous fission could be more consistently treated via explicitly considering the fission barrier transmission probabilities as was done in \cite{Samuel18} as compared to the simple parameterized form of spontaneous fission rates applied in this work. Additionally here we use FRDM2012 to describe less deformed nuclei and make use of FRLDM for our fission treatments since the former is more appropriate to describe ground state properties while the latter describes strongly deformed systems, however a unified single description applicable in both regimes would be ideal. Although such considerations are important goals of future efforts, achieving complete consistency is beyond the scope of present calculations. Thus this work is meant to highlight the recent steps toward consistency achieved via the application of FRLDM for both the fission yields and rates as well as to motivate progress in theoretical nuclear data treatments for the $r$ process.

We note that the the most important nuclei influencing our astrophysical arguments were not those found to undergo fission furthest from stability past $N=184$ since deposition from such nuclei re-equilibrates to the the $r$-process path at early times. Rather it is the late-time deposition from the fissioning nuclei along the route back to stability that builds up the light precious metal region and could be responsible for the co-production signature seen in the elemental ratios. The wide fission yields predicted for these nuclei by the FRLDM fission yield model that are responsible for such co-production behavior have now also been reported by recent microscopic calculations for two $r$-process nuclei, $^{254}$Pu and $^{260}$Fm \citep{Sadhukhan}. Should this behavior be found to be common by future independent theoretical fission yield calculations or confirmed via experiment, such findings could give further credence to the potential fission signatures discussed in this work.

\acknowledgments

N.V. would like to thank Hendrik Schatz, Erika Holmbeck, Jinmi Yoon, and Benoit C\^ot\'e for useful discussions as well as David Radice for providing simulation data. The work of N.V., G.C.M., M.R.M., and R.S. was partly supported by the Fission In R-process Elements (FIRE) topical collaboration in nuclear theory, funded by the U.S. Department of Energy. Additional support was provided by the U.S. Department of Energy through contract numbers DE-FG02-02ER41216 (G.C.M), DE-FG02-95-ER40934 (R.S. and T.M.S.), and DE-SC0018232 (SciDAC TEAMS collaboration, R.S. and T.M.S). R.S. and G.C.M also acknowledge support by the National Science Foundation Hub (N3AS) grant No. PHY-1630782.
M.R.M. was supported by the US Department of Energy through the Los Alamos National Laboratory. Los Alamos National Laboratory is operated by Triad National Security, LLC, for the National Nuclear Security Administration of U.S.\ Department of Energy (Contract No.\ 89233218CNA000001). 
This work was partially enabled by the National Science Foundation under grant No. PHY-1430152 (JINA Center for the Evolution of the Elements). 
This manuscript has been released via Los Alamos National Laboratory report number LA-UR-19-30652. 


%

\vspace{5mm}




\bibliographystyle{yahapj}
\bibliography{frldmRPrefs}



\end{document}